\documentclass[journal]{IEEEtran}

\ifCLASSINFOpdf
\else
\fi

\usepackage{array}
\usepackage{times}
\usepackage{epsfig}
\usepackage{graphicx}
\usepackage{amsmath}
\usepackage{amssymb}
\usepackage{graphicx}
\usepackage{cite}
\usepackage{enumerate}
\usepackage{cases}
\usepackage{multirow}
\usepackage{verbatim}
\usepackage{amssymb}
\usepackage{CJK}
\usepackage{algorithm}
\usepackage{algorithmicx}
\usepackage{algpseudocode}
\usepackage{stfloats}
\usepackage{color}

\usepackage{bm}
\usepackage{booktabs}
\usepackage[colorlinks,linkcolor=red]{hyperref}

\begin{document}

\title{Learning to Remove Clutter in 
Real-World GPR Images Using Hybrid Data}

% CYBERNETICS

\author{Hai-Han Sun, Weixia Cheng, and Zheng Fan
% <-this % stops a space%

}% <-this % stops a space}

% The paper headers
\markboth{}%
{Shell \MakeLowercase{\textit{et al.}}: Bare Demo of IEEEtran.cls for IEEE Journals}

% make the title area
\maketitle

% As a general rule, do not put math, special symbols or citations
% in the abstract or keywords.
\begin{abstract}
The clutter in the ground-penetrating radar (GPR) radargram disguises or distorts subsurface target responses, which severely affects the accuracy of target detection and identification. Existing clutter removal methods either leave residual clutter or deform target responses when facing complex and irregular clutter in the real-world radargram. To tackle the challenge of clutter removal in real scenarios, a clutter-removal neural network (CR-Net) trained on a large-scale hybrid dataset is presented in this study. The CR-Net integrates residual dense blocks into the U-Net architecture to enhance its capability in clutter suppression and target reflection restoration. The combination of the mean absolute error (MAE) loss and the multi-scale structural similarity (MS-SSIM) loss is used to effectively drive the optimization of the network. To train the proposed CR-Net to remove complex and diverse clutter in real-world radargrams, the first large-scale hybrid dataset named CLT-GPR dataset containing clutter collected by different GPR systems in multiple scenarios is built. The CLT-GPR dataset significantly improves the generalizability of the network to remove clutter in real-world GPR radargrams. Extensive experimental results demonstrate that the CR-Net achieves superior performance over existing methods in removing clutter and restoring target responses in diverse real-world scenarios. Moreover, the CR-Net with its end-to-end design does not require manual parameter tuning, making it highly suitable for automatically producing clutter-free radargrams in GPR applications. The CLT-GPR dataset and the code implemented in the paper can be found at https://haihan-sun.github.io/GPR.html. 

\end{abstract}

% Note that keywords are not normally used for peer review papers.

\IEEEpeerreviewmaketitle

\section{Introduction}

\IEEEPARstart {G}{round-penetrating} radar (GPR) has been a powerful tool for detecting subsurface targets in many fields, such as civil engineering, geology engineering, and military  \cite{b1,b2}. However, the strong background clutter in the GPR radargram caused by the reflection from the air-to-ground interface and the heterogeneity of subsurface materials disguises or obscures the target response, especially when the target is at a shallow depth. Wideband GPR systems have been implemented \cite{wb1,wb2,wb3,wb4,wb5,wb6} and high-resolution algorithms have been developed \cite{hr1,hr2,hr3,hr4,hr5,hr6,hr7} to enhance discrimination between targets and clutter. Since the presence of environmental clutter in the radargram significantly challenges the accuracy of target detection and interpretation, it is imperative to eliminate environmental clutter and restore target responses, which is the focus of this work.\par

Over the years, many methods have been investigated to suppress clutter in GPR radargrams. Conventional methods include mean subtraction methods \cite{ms1,ms2,ms3} and subspace-based methods such as singular value decomposition (SVD) \cite{svd1,svd2,svd3}, principal component analysis (PCA), and independent component analysis (ICA) \cite{ica1,ica2}. The mean subtraction methods regard the averaged trace of raw traces in a selected window in a B-scan as the background clutter trace, and subtract it from raw traces to eliminate clutter. They perform well in homogeneous subsurface environments with flat surfaces where the averaged trace in no-object regions fully captures the background clutter information. However, their performance is degraded in heterogeneous environments with rough surfaces where the background clutter varies from trace to trace in the B-scan \cite{svd3}. The subspace-based methods project the original B-scan matrix onto the clutter and target components based on  the  difference in signal strength, and remove the clutter components from the raw data to mitigate clutter. They assume that the clutter signal is much stronger than the target response and thus the clutter information is contained in the dominant components. However, the clutter and target components cannot be well separated when their signals have similar strengths. As a result, the target response could be deteriorated by the clutter removal operation using these methods \cite{mine1}. \par

Based on the prior that the clutter caused by antennas’ crosstalk and surface reflection is relatively constant so it has a low-rank characteristic, whereas the target response has a sparse property, non-negative matrix factorization (NMF) \cite{nmf} is proposed to extract the low-rank components and remove them from the raw data to suppress clutter. Following that, the methods based on low-rank and sparse decomposition (LRSD) are investigated, including the robust PCA (RPCA) \cite{rpca1,rpca2}, go decomposition (GoDec) \cite{godec1,godec2}, and robust NMF (RNMF) \cite{rnmf}. These methods separate the target response and the clutter by optimizing a low-rank and sparse matrix representation problem. However, the target response that has wide and flat hyperbolic features can degrade the sparsity and result in low effectiveness of these methods \cite{rpca2}. Moreover, the performance of these methods highly depends on the selection of hyperparameters that control the trade-off between the low-rank components and sparse components, and the suitable hyperparameters vary from case to case \cite{rnmf}. Therefore, these methods may have inconsistent clutter removal performance under different subsurface scenarios. Similarly, the morphological component analysis (MCA) \cite{mca} and dictionary learning-based method \cite{dictionary} are presented to separate the clutter and target responses using their respective sparse representations or dictionaries. The target responses are subsequently reconstructed using the target dictionaries. The clutter removal performance of these methods highly depends on the selection of accurate dictionaries \cite{cae}. Considering the limitations of the conventional methods, a more effective solution is desired to improve the clutter removal performance in diverse real-world subsurface environments.\par

With the strong feature representation and learning capability, deep learning-based methods have been involved in solving challenging GPR tasks, such as target detection \cite{tar1,tar2}, characterization \cite{characterization1,characterization2,characterization3}, and inverse imaging \cite{inverseimaging1,inverseimaging2}. There is a rising trend to apply neural networks in the clutter removal task. In \cite{rae}, a robust autoencoder-based method is introduced to solve the low-rank and sparse matrix representation problem. This method achieves better performance than RPCA \cite{rpca1} as the autoencoder offers non-linear solutions. In \cite{rnmf-net}, a network model constructed with 1D convolutional layers is presented to separate target responses from clutter in GPR A-scans. By using low-rank components and sparse components decomposed by RNMF as guidance, the network achieves similar performance with RNMF without tuning hyperparameters. In \cite{cae}, a network model based on the convolutional autoencoder (CAE) is presented to learn the clutter removal in 2D B-scans using a synthetic dataset. The network outperforms most of the existing methods in multiple scenarios, but its clutter removal capability is degraded in real-world GPR radargrams. This is because the network is only trained on the synthetic data that cannot capture the complex and diverse real-world clutter distributions.\par

Previous deep learning-based methods have shown promising capability in removing GPR clutter, but their generalizability for real-world scenarios is highly constrained by the synthetic training data and the simple network structure. This work devotes to address the challenges of deep learning-based methods in removing clutter in real-world scenarios while enhancing the clutter removal capability in general by making three contributions:\par

\begin{table*}[bp]
\caption{Properties of the Three Sub-Datasets in the CLT-GPR Dataset}
\centering
\begin{tabular}{c c c c}
\hline
\hline
\textbf{Sub-dataset}           & \textbf{GPR   system}                                                                                                                             & \textbf{Subsurface   environment}                                                                           & \textbf{Number of data} \\ \hline
\textbf{Synthetic sub-dataset} & 1.5-GHz   GSSI antenna in gprMax \cite{gprmax1,gprmax2,gprmaxant1,gprmaxant2}                                                                                               & \begin{tabular}[c]{@{}c@{}}Six   different soil conditions with \\ four types of soil surfaces\end{tabular} & 1,920                   \\
\hline
\textbf{Sand sub-dataset}      & Multi-polarimetric GPR system                                                                                                        & \begin{tabular}[c]{@{}c@{}}Sandy soil with uneven surface and \\ random distributed moisture content \end{tabular}                                                                                               & 6,000                   \\
\hline
\textbf{Concrete sub-dataset}        & \begin{tabular}[c]{@{}c@{}}Single-polarimetric GPR system with mono-static \\ and bi-static antenna configurations\end{tabular} & Concrete                                                                                                    & 4,000                   \\ 
\hline
\hline
\end{tabular}
\label{table:dataset}
\end{table*}

\begin{itemize}
\item In view of the limitations of using synthetic data to train the data-driven clutter removal method, we built the first large-scale hybrid dataset with real-world clutter distributions named CLT-GPR dataset. The proposed dataset contains diverse clutter distributions from both the synthetic data and measured data collected using different GPR systems in different environments. Experimental results verify that the dataset serving as training data could significantly boost the generalization capability of the data-driven method in eliminating clutter in real-world radargrams. 

\item Different from previous deep learning-based methods that use simple network architectures, we leverage the powerful learning capability of deep neural network and present a clutter-removal network (CR-Net). The CR-Net is featured by the integration of residual dense blocks (RDBs) into the U-Net architecture. The network extracts the hierarchical features of the input raw radargram with different receptive fields, retains the features related to the target reflection and suppresses the unwanted clutter features, and finally reconstructs the clutter-free radargram. The inclusion of RDBs enhances the network capability in adaptively removing clutter without deteriorating target responses. 

\item The loss function that drives network optimization has a critical impact on the network performance. In this work, instead of using conventionally adopted mean squared error (MSE) loss, we take one step forward by investigating the effects of different losses on the network performance for the clutter removal task. It is experimentally proved that the combined loss function of the mean absolute error (MAE) loss and the multi-scale structure similarity (MS-SSIM) loss leads to greatly improved clutter removal performance than the MSE loss. 

\end{itemize}

Extensive experiments demonstrate that the built CLT-GPR dataset, the design of the CR-Net, and the combined loss function benefit from each other, producing a network with superior and consistent performance in suppressing clutter in diverse scenarios. The well-trained CR-Net automatically removes clutter and restores the target response without any selection of hyperparameters. Moreover, several case studies verify that our method outperforms the existing methods by a large margin in terms of both quantitative metrics and visual quality. \par

The rest of the paper is arranged as follows. Section II describes the collection of the CLT-GPR dataset. Section III presents the CR-Net architecture and the combined loss function that drive the network optimization. In Section IV, extensive experiments are conducted to evaluate the performance of the CR-Net and demonstrate the advantages of our method in various scenarios. Finally, conclusion is drawn in Section V. \par

\section{CLT-GPR Dataset Collection}\label{sec2}

As the existing dataset for the clutter removal task only contains synthetic data \cite{cae,rnmf-net}, it cannot allow data-driven methods to learn complex and diverse clutter distributions in real-world radargrams, thus constraining the generalization capability of the data-driven methods in real scenarios. To address this issue, we built the CLT-GPR dataset containing diverse real-world clutter for training data-driven clutter removal methods. The CLT-GPR dataset is composed of three sub-datasets: a synthetic sub-dataset, a sand sub-dataset, and a concrete sub-dataset. The characteristics of each sub-dataset are listed in Table \ref{table:dataset}. The synthetic sub-dataset includes simulated clutter in different soil and surface conditions, which is detailed in subsection A. The sand sub-dataset includes clutter collected by a multi-polarimetric GPR system in a sandy field, which is described in subsection B. The concrete sub-dataset contains clutter collected by a single-polarized GPR system with mono-static and bi-static antenna configurations in concrete environment, which is presented in subsection C. The different GPR systems and subsurface scenarios in the dataset preparation provide diverse distributions of the real-world clutter. Each pair of data in the dataset includes a raw B-scan radargram and its corresponding clutter-free image, which facilitates the supervised learning of data-driven methods. The preparation details of these sub-datasets are described as follows.\par

\subsection{Synthetic Sub-Dataset}
\begin{figure}[tb]
	\begin{center}
		\begin{tabular}{c@{ }}
			\includegraphics[width=0.98\linewidth]{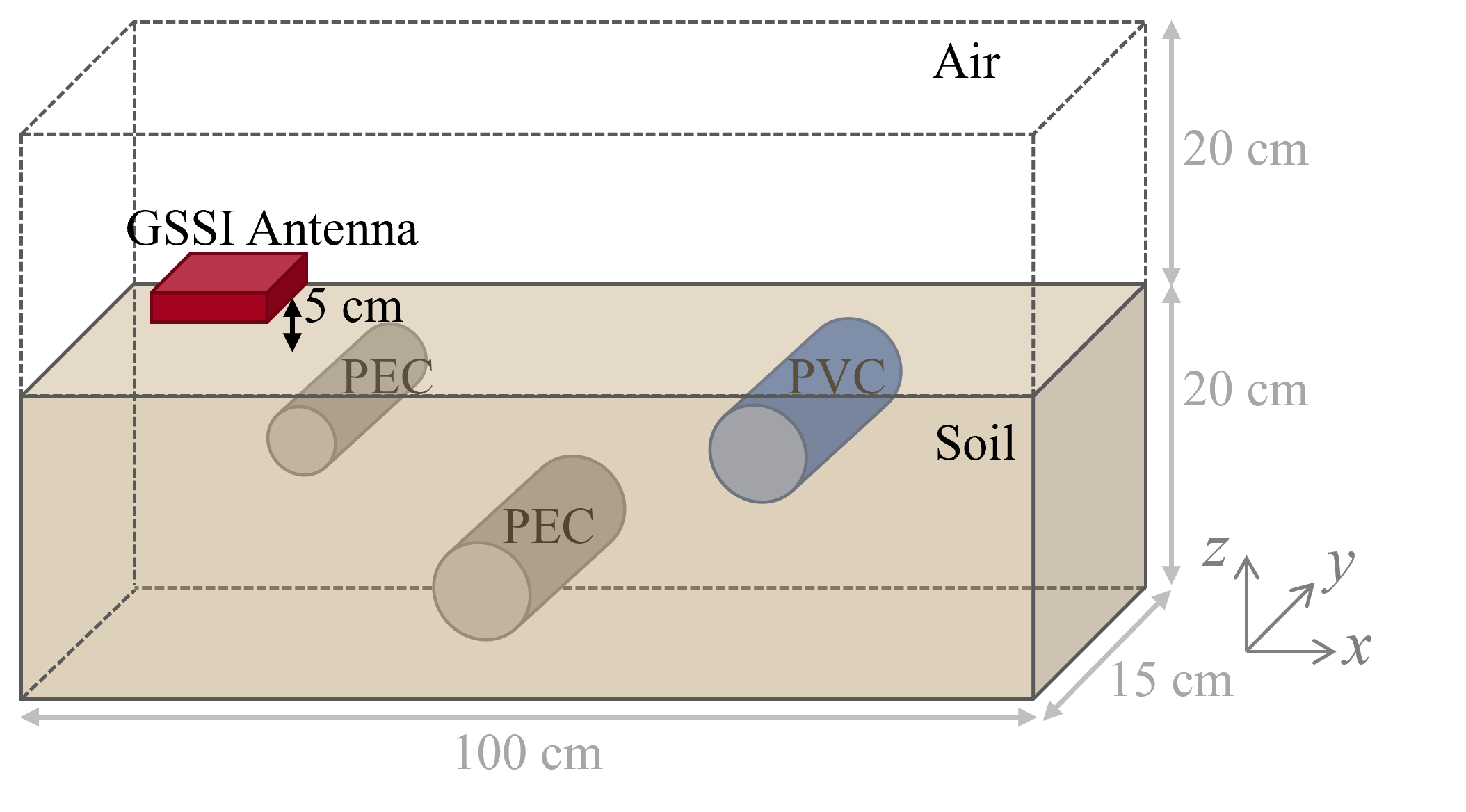}\\
			\footnotesize{(a)} \\
			\includegraphics[width=0.98\linewidth]{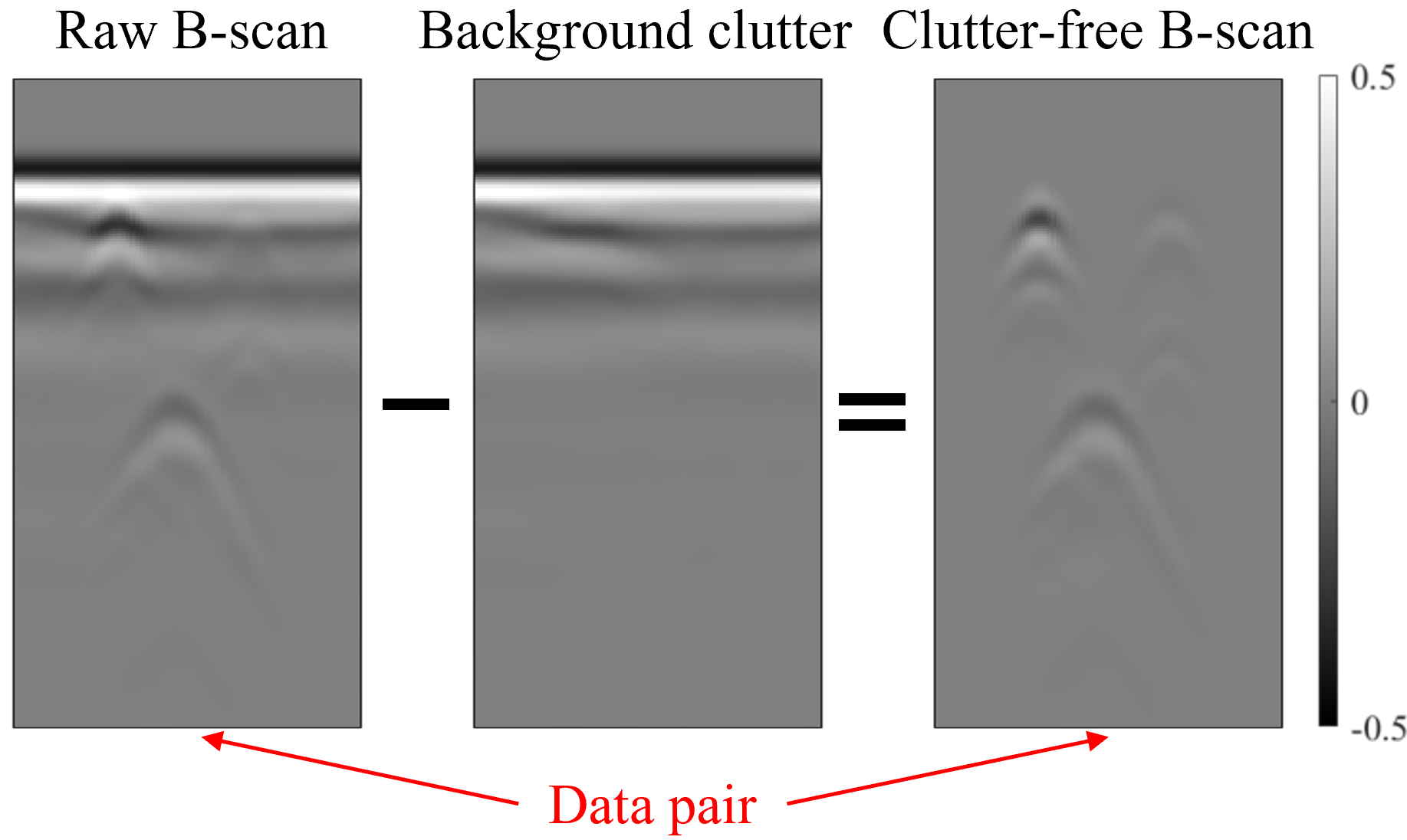}\\
			\footnotesize{(b)}  \\
		\end{tabular}
	\end{center}
	\caption{(a) Illustration of the simulation scenario for the preparation of the synthetic sub-dataset. The 1.5-GHz GSSI antenna is used as the transmitter and receiver to collect GPR signals. Four types of soil surfaces and six types of soil are included as soil conditions to generate diverse subsurface environments. Two types of cylindrical objects (i.e., PVC and PEC) with different radii are included as subsurface targets, and they are placed at different positions. (b) Illustration of the raw B-scan, the corresponding background clutter, and the clutter-free B-scan. The clutter-free B-scan is obtained by subtracting the background clutter from the raw B-scan.}
	\label{fig:1}
\end{figure}

A synthetic GPR dataset is generated using the GPU-accelerated open-source software gprMax \cite{gprmax1,gprmax2}. The simulation scenario is shown in Fig. \ref{fig:1}(a). The simulation domain covers an area of 100×15×40 cm$^3$. The absorbing boundary condition is applied to reduce the side reflection. The spatial discretization is 0.2 cm in the $x$-, $y$- and $z$-directions. The build-in GSSI antenna operating at 1.5 GHz is used as the transmitter and receiver \cite{gprmaxant1,gprmaxant2}. It has a size of 17×10.8×4.5 cm$^3$ and is located 5 cm above the soil surface. The antenna is moved along the $x$-direction with a step of 1 cm. 80 A-scans are collected along the scan traces to generate a B-scan. \par

To generate a synthetic dataset covering diverse subsurface scenarios, four types of surfaces, six types of soil, two types of objects, and three numbers of subsurface objects with different depths and radii are modeled in the simulation.\par

The types of surfaces cover common scenarios, including the flat surface, the grass surface, the rough surface, and the rough surface with surface water. For the grass surface, 10,000 blades of grass are randomly distributed on a flat surface. The height of the blades of grass and the depth of the grass roots both vary from 0 to 4 cm. For the rough surface, the roughness is randomly distributed with a fluctuation range of 4 cm. For the rough surface with surface water, the surface roughness is the same as the rough surface cases, but surface water is added with a depth of 1 cm. \par

The six types of soil include dry sand, damp sand, dry clay soil, wet clay soil, dry loam soil, and heterogeneous soil. The material properties of different soil conditions (except the heterogeneous soil) are listed in Table \ref{table:material}. The heterogeneous soil is composed of 50 different soil fractures with a sand fraction of 0.5, a clay fraction of 0.5, a soil bulk density of 2 g/cm$^3$, a sand density of 2.66 g/cm$^3$, a water volumetric fraction of 0.1\%-25\%, and fractal dimension of 1.5. \par

The objects are cylinders made of the perfect electric conductor (PEC) or polyvinyl chloride (PVC). The material properties of the modeled PVC are also listed in Table \ref{table:material}. In each combination of soil and surface scenario, cylindrical objects with their radii vary from 1 cm to 5 cm are buried along the $y$-direction. The number of subsurface objects varies from one to three. For the one-object case, 20 B-scans are generated, in which 10 B-scans for PVC objects and 10 B-scans for PEC objects. The object is located at the horizontal position of 50 cm but with depths varying from 1 cm to 10 cm. For the two-object and three-object cases, 30 B-scans are generated for each case. The materials of the objects are randomly selected from PEC and PVC, and the horizontal positions and cover depths are randomly selected in the range of 20-80 cm and 0-10 cm, respectively. Special care is taken to avoid the overlapping of multiple objects. A total of 80 B-scans are generated for each combination of soil and surface scenario.\par

\begin{table}[tb]
	\caption{Electromagnetic Properties of the Materials for the Synthetic Sub-Dataset}
	\centering
	\begin{tabular}{c c c}
		\hline
		\hline
		\textbf{Material}   & \textbf{Relative permittivity } & \textbf{Conductivity (S/m)}  \\
		\hline
\textbf{Dry sand}  &  3.0     &	 0.001\\
\textbf{Damp sand}  &  8.0     &	 0.01\\
\textbf{Dry clay soil}  &  10.0     &	0.01\\
\textbf{Wet clay soil}  &  12.0     &	 0.01\\
\textbf{Dry loam soil}  &  10.0     &    0.001\\
\textbf{PVC}  &  3.5     &	 0\\
		\hline
		\hline
	\end{tabular}
	\label{table:material}
\end{table}

In total, four types of surfaces, six types of soil, and 80 different object arrangements in each soil and surface scenario produces 1,920 raw B-scans. Their corresponding clutter-only background B-scans are obtained by maintaining the same soil condition but without burying objects. The clutter-free B-scans are obtained by subtracting the background B-scans from the raw B-scans. An example of the raw, background, and clutter-free B-scans of a three-object case is shown in Fig. \ref{fig:1}(b) to illustrate the process of obtaining a pair of data.

\subsection{Sand Sub-Dataset}
As the ideal clutter-free image cannot be obtained from the raw radargram in the field experiment, we adopt an alternative method to generate a pair of raw B-scan and its corresponding clutter-free B-scan. We first collect clutter-only radargrams in a controlled environment, and then combine them with the simulated clutter-free images as presented in Section II.A to produce hybrid raw GPR images. These hybrid raw images and their corresponding clutter-free images form the experimental dataset. \par

\begin{figure}[tb]
	\begin{center}
		\begin{tabular}{c@{ }}
			\includegraphics[width=0.98\linewidth]{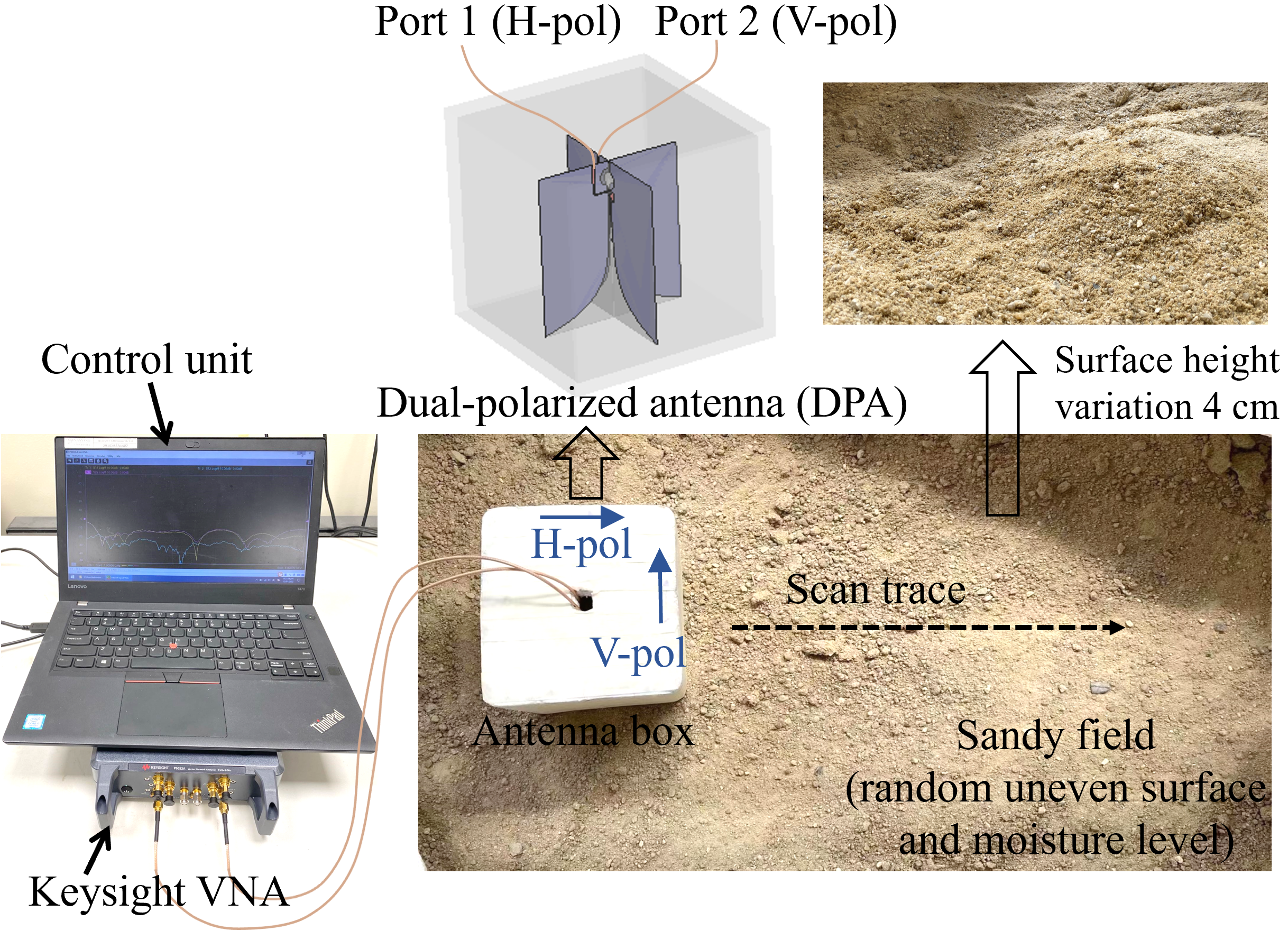}\\
			\footnotesize{(a)} \\
			\includegraphics[width=0.98\linewidth]{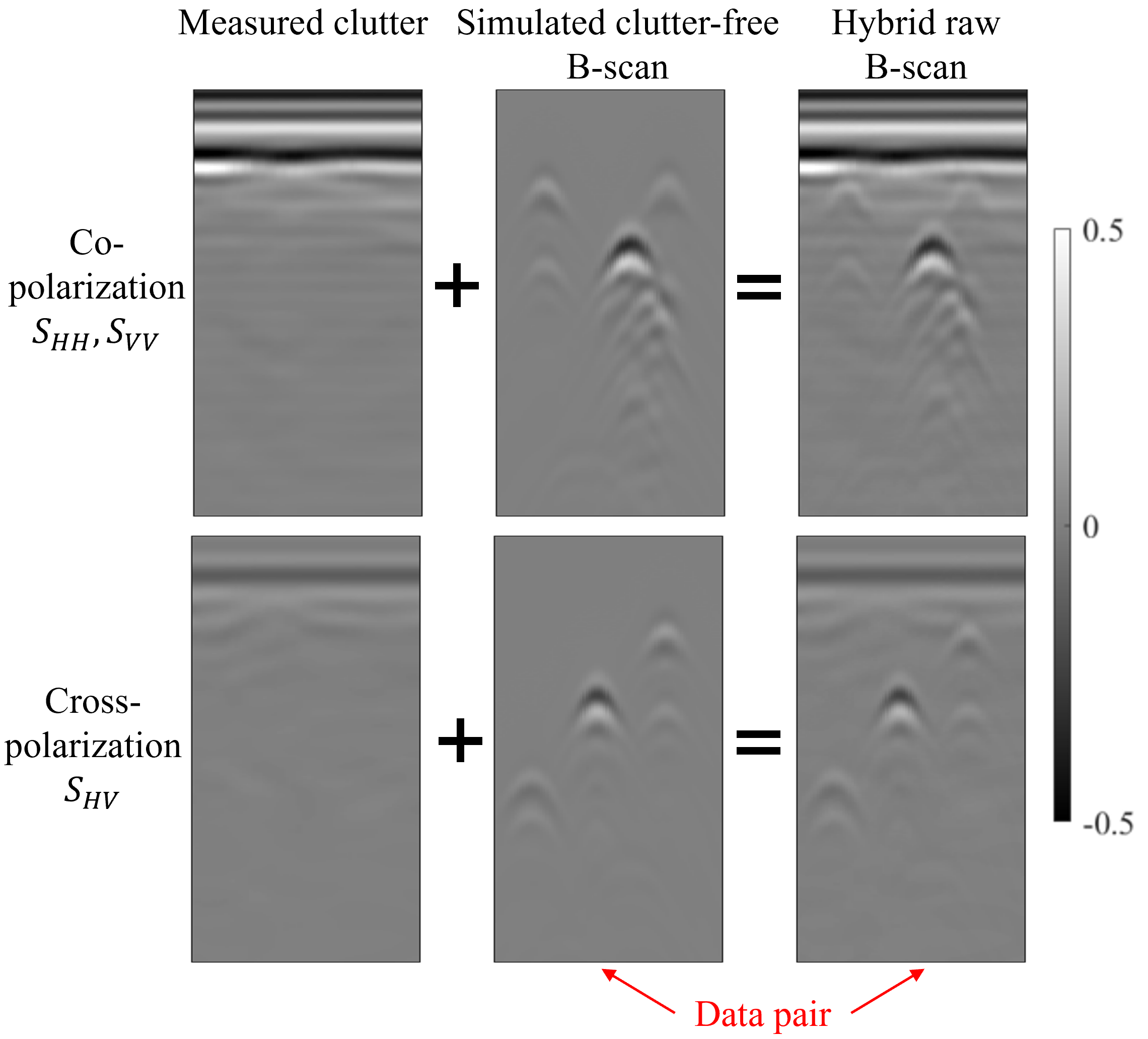}\\
			\footnotesize{(b)}  \\
		\end{tabular}
	\end{center}
	\caption{(a) Demonstration of the measurement scenario using a multi-polarimetric GPR system to collect real clutter data in a controlled sandy field without any subsurface objects. The sandy field has an uneven surface and random distributed moisture levels. (b) Illustration of the collected clutter-only radargram of the co-polarized components ($S_{HH}$ and $S_{VV}$), and the cross-polarized components ($S_{HV}$), and the process to generate the hybrid B-scan by combining the measured clutter with a simulated clutter-free radargram.}
	\label{fig:DPA}
\end{figure}

In the first field experiment, a multi-polarimetric GPR system is employed to collect real clutter in sandy soil without any subsurface objects, as shown in Fig. \ref{fig:DPA}(a). The sandy field has an uneven surface and randomly distributed moisture levels in the measurement. The height variation of the soil surface is within 4 cm. Different moisture levels are created by adding water to different portions of the soil. The resultant relative permittivity of the soil varies from 3 to 8. These settings are created to increase clutter diversity.  The dual-polarized Vivaldi antennas (DPA) operating from 1.3 GHz to 6.0 GHz are used as the transmitter and receiver in a mono-static setup. The antenna is sealed in a foam box and is surrounded by absorbers to reduce noise from the surrounding environment. The antenna has two ports 1 and 2 that excite the horizontal-polarized (H-pol) and vertical-polarized (V-pol) radiation, respectively. The two ports are connected to a vector network analyzer (Keysight VNA P5022A), forming a multi-polarimetric stepped-frequency GPR system. The system simultaneously collects three A-scans of the three polarimetric components, including co-polarized components $S_{HH}$ and $S_{VV}$ and cross-polarized component $S_{HV}$. The first and second subscripts of the scattering components denote the polarization of receiving and transmitting antennas. For each A-scan, 1,001 samples in a frequency band from 1.3 GHz to 6.0 GHz are recorded. The collected frequency domain data are transformed to the time domain via inverse Fourier transform. For each polarimetric component, 64 A-scans are collected along a scan trace with a step size of 1 cm and are combined into a clutter-only B-scan. \par

In the experiment, 300 clutter-only B-scans of different polarimetric components are collected. To further increase the clutter diversity, four different frequency bands are selected to process the acquired frequency domain data, which are 1.3 - 3.3 GHz, 1.3 - 4.3 GHz, 1.3 - 5.3 GHz, and 1.3 - 6.0 GHz, respectively. A total of 1,200 clutter-only B-scans are obtained. Each radargram is resized to 256×64 and combined with five randomly selected simulated clutter-free radargrams, forming 6,000 hybrid data. The clutter-free images are also resized to 256×64 to ease the combination process. The hybrid data and their corresponding clutter-free images form the sand sub-dataset. It is noted that the simulated target response in the clutter-free image was obtained using the 1.5 GHz GPR and therefore may not exactly match the signal response in other frequency bands. However, the objective of the hybrid dataset is to allow the data-driven methods to capture the features of real clutter and remove them, where the presence of the target response only provides supplementary help for the methods to distinguish the characteristics of the clutter from the target response. The operation to produce the hybrid dataset is feasible for the clutter removal task, as demonstrated by subsequent experiments with the dataset. \par

Fig. \ref{fig:DPA}(b) illustrates the process of generating the sand sub-dataset. As the soil surface is uneven and the sandy field is heterogeneous, unevenness of the ground reflection and subsurface environmental clutter can be observed in the collected clutter radargram.  In addition, it can be observed that the clutter collected by the co-polarized component and the cross-polarized component are significantly different. The cross-polarized component receives much less antenna crosstalk and surface reflection clutter than the co-polarized components due to polarization mismatch \cite{xpol}. The clutter collected by different polarimetric components of the GPR system further increases the clutter diversity in the sand sub-dataset. After combining the clutter radargram with the simulated clutter-free radargram, the background clutter interferes and disguises the target responses, which imitates the radargram collected in real scenarios. Therefore, the obtained hybrid radargram contains diverse clutter distributions and resembles the radargrams in real cases. \par

\begin{figure}[tb]
	\begin{center}
		\begin{tabular}{c@{ }}
			\includegraphics[width=0.98\linewidth]{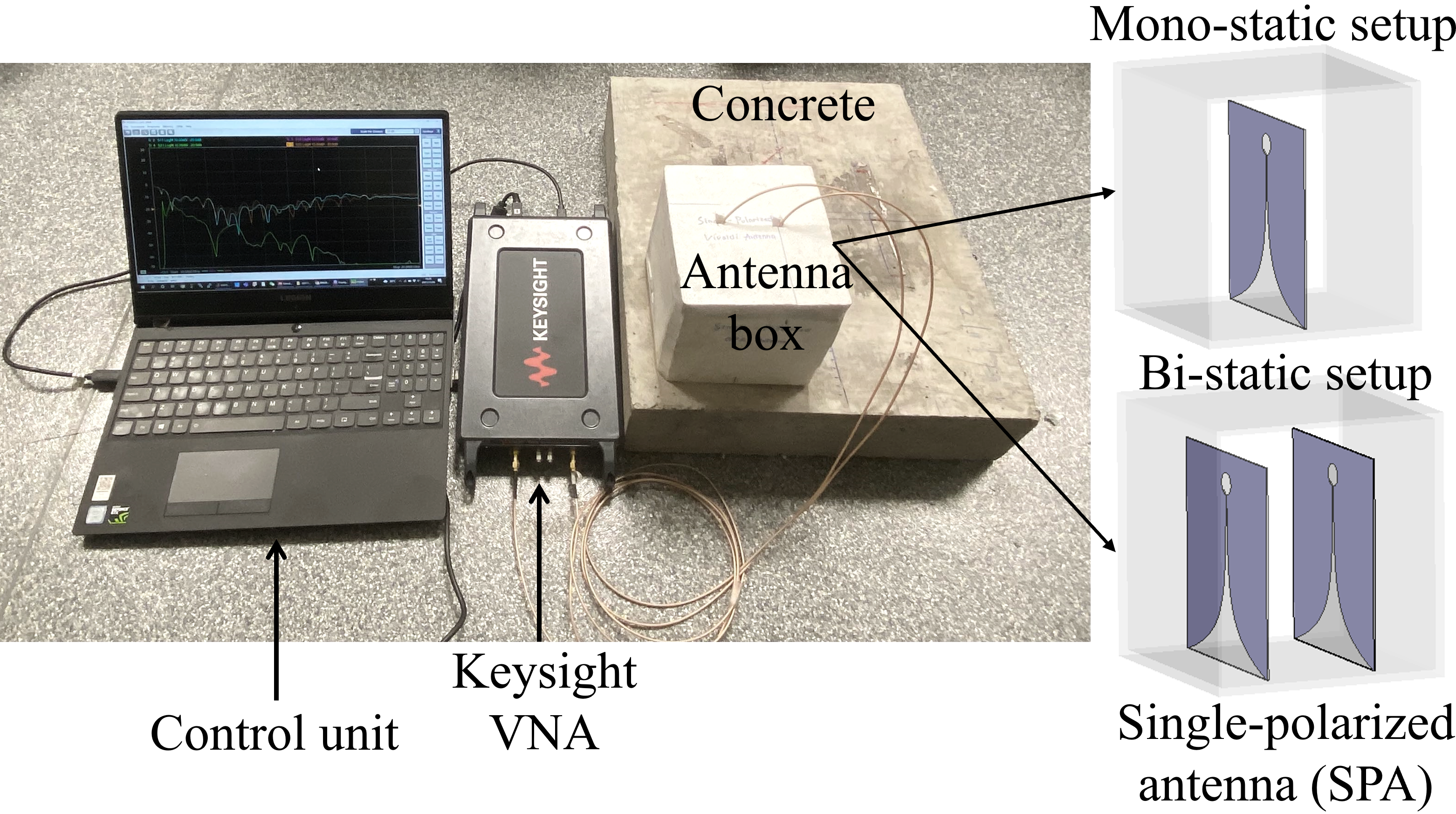}\\
			\footnotesize{(a)} \\
			\includegraphics[width=0.98\linewidth]{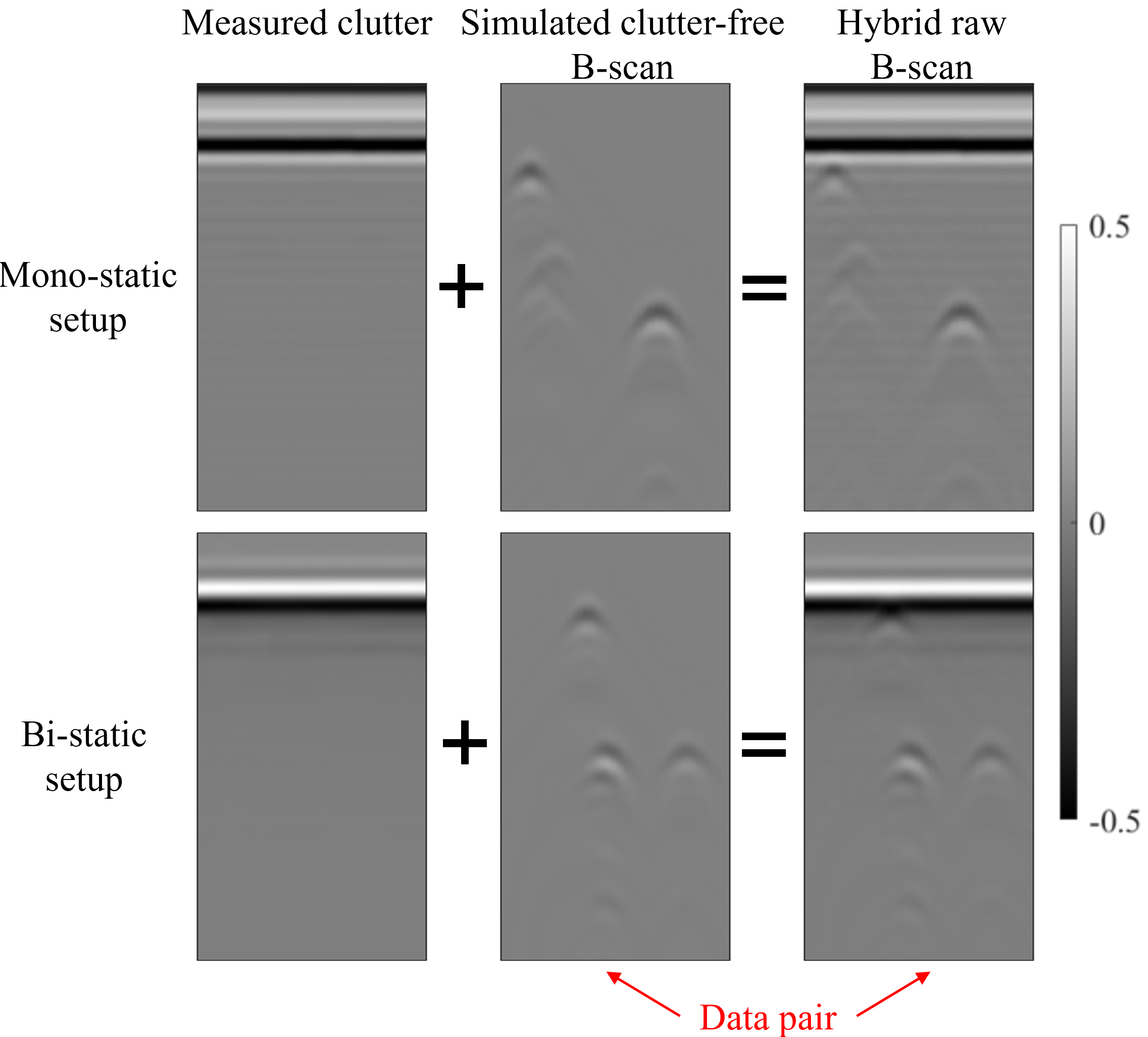}\\
			\footnotesize{(b)}  \\
		\end{tabular}
	\end{center}
	\caption{(a) Demonstration of the measurement scenario using a stepped-frequency GPR system with single-polarized Vivaldi antenna (SPA) in mono-static and bi-static setups to collect real clutter data. The data are collected on rebar-free concrete samples. (b) Illustration of the collected clutter radargrams by the mono-static and bi-static setups and the process to obtain hybrid B-scans. }
	\label{spa}
\end{figure}

\subsection{Concrete Sub-Dataset}

In the second field experiment, a single-polarimetric GPR system with both mono-static and bi-static antenna configurations is used to collect clutter data in concrete environments, as shown in Fig. \ref{spa}(a). The system setup is similar to the one described in Section II.B, except that the equipped antenna is the single-polarized Vivaldi antenna (SPA). Both the mono-static and bi-static antenna configurations are implemented as they provide different clutter coming from the antenna and surface reflection, which enhances the clutter diversity. For each A-scan, the GPR system records 1001 frequency samples from 1.3 GHz to 6.0 GHz. 64 A-scans are collected and are combined into a B-scan.  \par

In the experiment, 200 clutter-only B-scans are collected on 15 concrete samples. Four different frequency bands are selected to process the collected data to further increase the clutter diversity, which are 1.3 - 3.3 GHz, 1.3 - 4.3 GHz, 1.3 - 5.3 GHz, and 1.3 - 6.0 GHz, respectively. The data are subsequently converted into the time domain via Fourier transform. In total, 800 clutter-only B-scans are obtained. Each radargram is resized and combined with five randomly selected clutter-free radargrams, forming 4,000 hybrid data. The hybrid data and their corresponding clutter-free radargrams form the concrete sub-dataset.\par

Fig. \ref{spa}(b) shows examples of clutter radargrams collected by the mono-static and bi-static antenna configurations and the obtained hybrid images in the concrete environment. The mono-static setup produces more clutter than the bi-static setup due to antenna self-reflection and ringing effect. As the concrete has a relatively flat surface, the surface clutter presents as horizontal bandings. The surface clutter could severely disguise the target signal in the hybrid data, especially for shallowly located objects, which imitates radargrams of concrete buildings with shallowly buried rebars. \par

The CLT-GPR dataset consisting of the synthetic sub-dataset and the two hybrid sub-datasets contains diverse clutter distributions of different GPR systems in multiple scenarios. Therefore, such a large-scale and diverse dataset could provide powerful support for training deep neural networks to remove clutter and restore target responses in real-world scenes. We believe the release of the CLT-GPR dataset could facilitate the development of novel clutter removal methods. \par

\section{Clutter-Removal Neural Network }\label{sec3}

With the collection of CLT-GPR dataset, a clutter-removal neural network, named CR-Net, is designed to eliminate clutter while restoring target responses in radargrams. The network structure and the loss function are described as follows. \par

\begin{figure*}[t]
	\centering
	\centerline{\includegraphics[width = 1\linewidth]{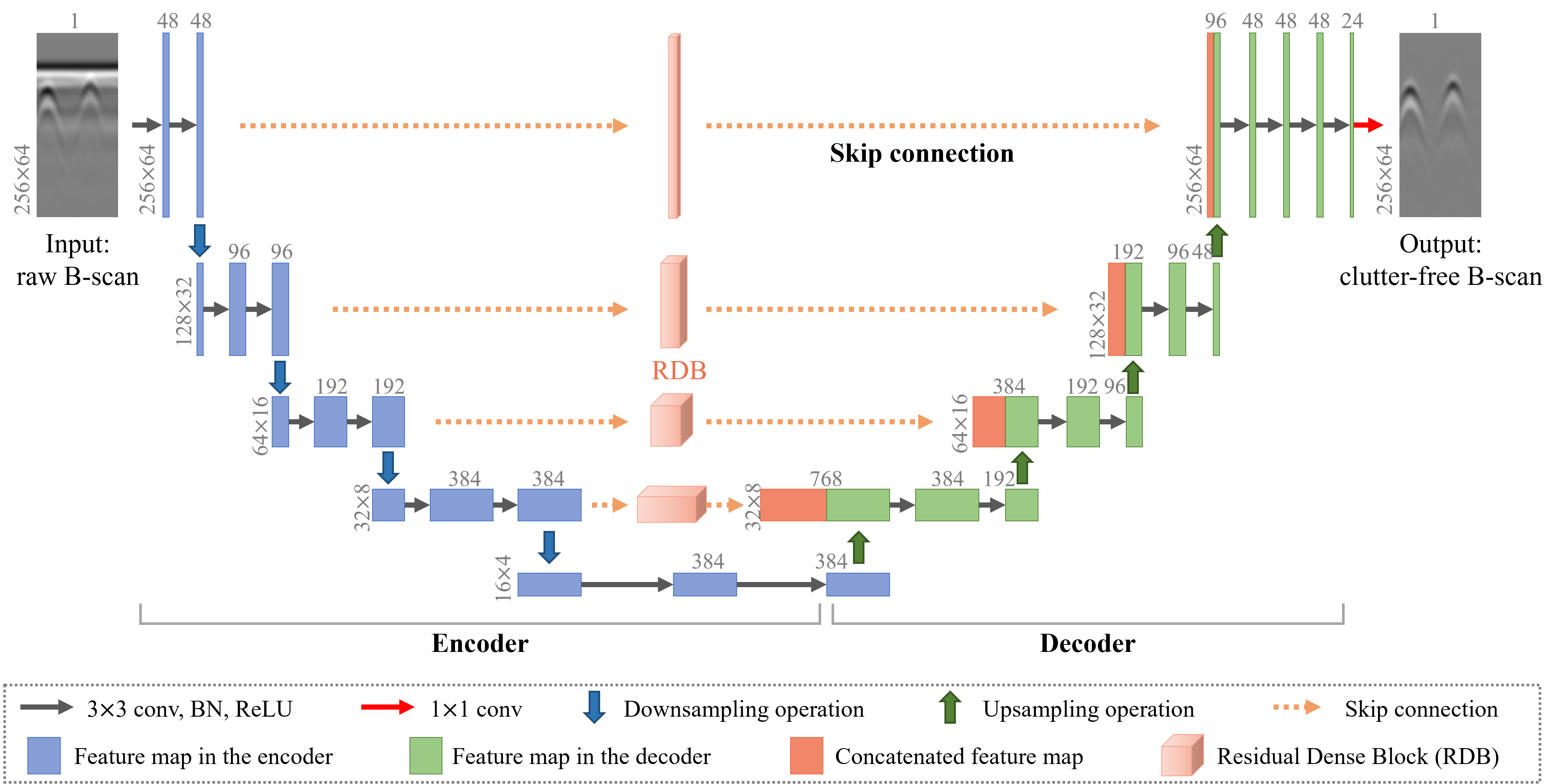}}
	\caption{Overview framework of the proposed CR-Net. It is composed of an encoder, a decoder, and skip connections between them. Residual dense blocks (RDBs) are integrated in the skip connection paths to adaptively preserve features related to object reflection and suppress features of clutter. The number of channels and the size of a feature map are denoted on the top and left of the box that representing a feature map, respectively. }
	\label{fig:net}
\end{figure*}

\begin{figure}[!t]
	\centering
	\centerline{\includegraphics[width=1\linewidth]{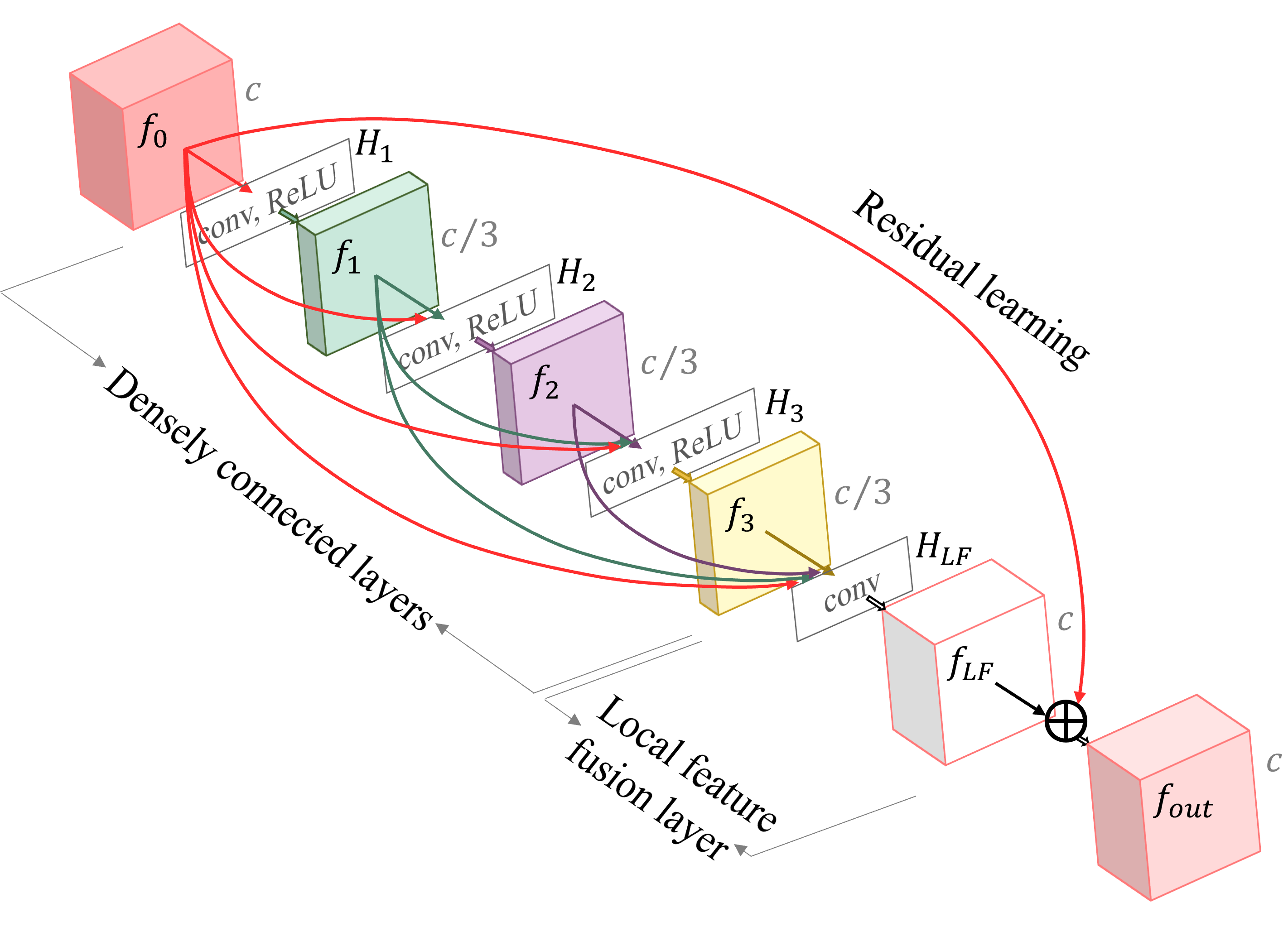}}
	\caption{Illustration of the residual dense block. It includes three densely connected layers, one local feature fusion layer, and one residual learning operation. The main purpose of the block is to adaptively preserve features related to target reflection and reduce the clutter features. $f$, $c$, and $H$ denote the feature map, the channel number of a feature map, and the non-linear transformation of different layers, respectively.}
	\label{fig:rdb}
\end{figure}

\subsection{Network Architecture}

The network architecture of the CR-Net is illustrated in Fig. \ref{fig:net}. It consists of an encoder, a decoder, and skip connections between them. The RDBs \cite{rdb} are integrated into the skip connection paths to adaptively preserve the target response information and reduce the clutter features.\par

The encoder is composed of four successive applications of two 3×3 convolutional layers and one downsampling operation. Each convolutional layer has a kernel size of 3×3, a stride of 1, and a padding of 1, and is followed by the batch normalization operation \cite{bn} and the rectified linear (ReLU) activation function\cite{relu}. The batch normalization is utilized to speed up training process and improve network performance \cite{bn}. The downsampling operation is implemented by a max-pooling layer with a kernel size of 2×2 and a stride of 2. The encoder with multi-scale layers extracts the hierarchical features of the input raw B-scan with different receptive fields.\par

The decoder includes four successive applications of one upsampling operation, a concatenation with the corresponding feature map from the encoder, and two 3×3 convolutional layers. The upsampling operation is implemented by the bilinear interpolation with a scale factor of 2. The convolutional layers are followed by the batch normalization operation and the ReLU activation function. \par

Instead of directly connecting the encoder features and decoder features that aims at avoiding the loss of target response information due to the downsampling operation in the encoder \cite{unet}, the RDBs are introduced between the encoder and decoder. It is found that the feature maps in the encoder, especially the low-level feature maps, include clutter information that has not been fully eliminated. Directly concatenating these encoder features to the decoder features could bring unwanted clutter information into the decoder, thus resulting in unsatisfactory clutter removal performance. Therefore, the feature maps in the encoder are first passed through a RDB to adaptively preserve the features related to the target response and reduce the unwanted clutter features, and then forwarded to the decoder. \par

The framework of the RDB is illustrated in Fig. \ref{fig:rdb}. It includes three densely connected layers, one local feature fusion layer, and one residual learning operation \cite{rdb}. In the densely connected layer, let $f_0$ and $f_l$ denote the input feature map and the output feature map of the $l^{th}$ layer, respectively, the input of the $l^{th}$ layer is the feature maps of all preceding layers $f_0$, $f_1$, …, $f_{l-1}$. The non-linear transformation of the $l^{th}$ layer $H_l (\cdot)$ is implemented by a 3×3 convolutional layer followed by the ReLU. The operation in the $l^{th}$ layer can be expressed as:
\begin{equation}
\label{equ_1}
f_l=H_l\left(\left[f_0,f_1,\ldots,f_{l-1}\right]\right),
\end{equation}
where [$\cdot$] is the concatenation of the feature maps along the channel dimension. Given that $f_0$ has $c$ channels and there are three densely connected layers, the channel number of $f_1$, $f_2$, and $f_3$ is set to $c⁄3$, which is referred as growth rate of RDB. The local feature fusion layer uses a 1×1 convolutional layer to adaptively fuse the outputs of the densely connected layers with the input feature map. The operation is formulated as:
\begin{equation}
\label{equ_2}
f_{LF}=H_{LF}\left(\left[f_0,f_1,f_2,f_3\right]\right),
\end{equation}
where $H_{LF}$ denotes a 1×1 convolution operation. The output channel number of $f_{LF}$ is reduced to $c$. Residual learning is introduced in the last stage to combine the fused feature map with the input feature map via channel-wise addition. The final output of the RDB is obtained as:
\begin{equation}
\label{equ_3}
f_{out}=f_0+f_{LF}.
\end{equation}
The RDB adaptively preserves the object reflection features while suppressing the clutter features in the feature maps from the encoder, which improves the network’s capability in removing background clutter while restoring target responses. \par

In the final stage of the CR-Net, two 3×3 convolutional layers and one 1×1 convolutional layer are employed to estimate the clutter-free B-scan. The number of channels and the size of all feature maps in the network are denoted in Fig. \ref{fig:net}. \par

\subsection{Loss Function}
To train the CR-Net, a loss function combining the MAE loss and the MS-SSIM loss \cite{ms-ssim,lossfunct} is used, which can be expressed as:
\begin{equation}
\label{equ_4}
{\rm L}={\rm MAE}\left(y,\hat{y}\right)+{\rm L_{MS\text{-SSIM}}}\left(y,\hat{y}\right).
\end{equation}
The loss function measures the dissimilarity between the predicted clutter-free image of the network $y$ and the ground-truth image $\hat{y}$. \par

The MAE loss is calculated as: 
\begin{equation}
\label{equ_5}
{\rm MAE}\left(y,\hat{y}\right)=\frac{1}{H\times W}\sum_{i,j}\left|y_{i,j}-{\hat{y}}_{i,j}\right|,
\end{equation}
where $H \times W$ is the dimension of the image, and $i$ and $j$ are the indices for a pixel value. \par

The MS-SSIM measures the perceived image quality of $y$ compared to its ground-truth image $\hat{y}$ in $M$ scales \cite{ms-ssim}. The image at the scale $k\in[1,M]$ is obtained by $(k-1)$ iterations of a low-pass filter and a down-sampling operation with a factor of 2. The MS-SSIM is defined as:
\begin{equation}
\label{equ_6}
{\rm MS\text{-SSIM}}\left(y,\hat{y}\right)=\left[l_M\left(y,\hat{y}\right)\right]^{\alpha_M}\cdot\prod_{k=1}^{M}{\left[c_k\left(y,\hat{y}\right)\right]^{\beta_k}\left[s_k\left(y,\hat{y}\right)\right]^{\gamma_k}}.
\end{equation}
In Eq. (6), $l\left(y,\hat{y}\right)$,$\ c\left(y,\hat{y}\right)$, and $s\left(y,\hat{y}\right)$ are the measures of luminance, contrast, and structure comparison given by
\begin{equation}
\label{equ_7}
l\left(y,\hat{y}\right)=\frac{2\mu_y\mu_{\hat{y}}+C_1}{\mu_y^2+\mu_{\hat{y}}^2+C_1},
\end{equation}
\begin{equation}
\label{equ_8}
c\left(y,\hat{y}\right)=\frac{2\sigma_y\sigma_{\hat{y}}+C_2}{{\sigma_y}^2+{\sigma_{\hat{y}}}^2+C_2},
\end{equation}
\begin{equation}
\label{equ_9}
s\left(y,\hat{y}\right)=\frac{\sigma_{y\hat{y}}+C_3}{\sigma_y\sigma_{\hat{y}}+C_3},
\end{equation}
where $\mu_y$ and $\mu_{\hat{y}}$ are the means of $y$ and $\hat{y}$, $\sigma_y$ and $\sigma_{\hat{y}}$ are the variances of $y$ and $\hat{y}$, respectively. $\sigma_{y\hat{y}}$ is the covariance of $y$ and $\hat{y}$. $C_1=\left(k_1R\right)^2$, $C_2=\left(k_2R\right)^2$, and $C_3=C_2/2$ are small constants, and $R$ is the dynamic range of the pixel values in the image. The subscript $k$ of the luminance, contrast, and structure comparison measures in Eq. (6) denotes the assessment at the $k$ scale. The exponents $\alpha_M$, $\beta_k$, and $\gamma_k$ are used to adjust the relative importance of different components at different scales. The values are calibrated and provided in \cite{ms-ssim}. The MS-SSIM loss is calculated as
\begin{equation}
\label{equ_10}
{\rm L_{MS\text{-SSIM}}}\left(y,\hat{y}\right)=1-{\rm MS\text{-SSIM}}\left(y,\hat{y}\right).
\end{equation}
\par
The utilization of MAE loss yields better convergence compared to using conventionally adopted MSE loss, and the MS-SSIM loss improves the visual similarity between the predicted image and the ground truth. The combined loss function in Eq. (4) is experimentally verified to strengthen the network’s clutter removal capability under the same training dataset. The comparison of network performance with different losses will be presented in Section IV.C. \par

\section{Experiments}\label{sec4}
In this section, the implementation details of the CR-Net are described in subsection A. The clutter removal performance of the CR-Net on the simulated and measured radargrams in multiple scenarios is presented in subsection B. The comparison of the clutter removal performance with existing methods is also provided. Subsection C presents the comparison of the clutter removal performance of different network architectures.  Ablation study is performed in subsection D to demonstrate the effects of the inclusion of hybrid sub-datasets into the training data, the integration of the RDB in the network architecture, and the use of different losses. The generalization capability of the CR-Net in a wider range of scenarios is examined in subsection E.  \par

\subsection{Implementation Details of the CR-Net}
The 1,920 sets of synthetic data as mentioned in Section II.A are split into training and testing datasets in an 80\%:20\% ratio. The 1,536 sets of synthetic training data are augmented to 4,608 sets by selected [1, 64], [9, 72], and [13, 76] ranges of A-scans as B-scan radargrams. The augmented synthetic training data are combined with the two hybrid sub-datasets as the final training data, which includes 14,608 raw B-scans with their corresponding clutter-free ground-truth images. All the radargrams are resized to 256×64 and normalized into a range of [0, 1]. \par
The 384 synthetic testing data, together with raw radargrams collected in different environments and radargrams in open-source GPR datasets are used to evaluate the clutter removal performance of the CR-Net, which is presented in subsection B.\par
The CR-Net is implemented with PyTorch on an NVIDIA 2080Ti GPU. The weights are initialized with the standard Gaussian function. The mini-batch size is set to 40. End-to-end training is performed based on the loss function defined in Eq. (4). The optimization of the network is performed using the ADAM optimizer \cite{adam} with default parameters. The initial learning rate is set to 0.0001 and decreases by a factor of 10 every 30 iterations. The model is trained for 100 epochs from scratch. \par

\subsection{Experimental Results }
The performance of the well-trained CR-Net is examined on the synthetic testing data and the measured testing data. Moreover, the CR-Net is implemented to remove clutter on radargrams in open-source GPR datasets to validate its generalization capability in real-world scenes. The performance of the subspace-based method SVD \cite{svd1,svd2,svd3}, the LRSD method RPCA \cite{rpca1}, and the deep-learning based CAE \cite{cae} is presented for comparison. The SVD is implemented by removing the largest singular value of the B-scan matrix. The hyperparameter $\lambda$ in RPCA is selected as 3×10$^{-2}$. The CAE model is built and implemented following the network design in \cite{cae}, and it is trained on the same training dataset as the proposed CR-Net.

\subsubsection{Results on the Simulated Testing Data}
As the simulated testing data have corresponding clutter-free images as ground truth, the clutter removal performance of different methods is evaluated using four full-reference-based metrics, including the MAE defined in Eq. (5), the MS-SSIM defined in Eq. (6), the MSE, and the peak signal-to-noise ratio (RSNR). The latter two metrics are defined as 
\begin{equation}
\label{equ_11}
{\rm MSE}=\frac{1}{H\times W}\sum_{i,j}\left(y_{i,j}-{\hat{y}}_{i,j}\right)^2,
\end{equation}
\begin{equation}
\label{equ_12}
{\rm PSNR}\left({\rm dB}\right)=10\cdot\log_{10}{\left(\frac{1}{{\rm MSE}}\right)}.
\end{equation}
In the Eqs. (11) and (12), $y$ and $\hat{y}$ are the normalized predicted clutter-free B-scan and the ground truth with a dimension of $H\times W$, and $i$ and $j$ are the indices for a pixel value. Smaller MAE and MSE values and larger PSNR and MS-SSIM values suggest a higher similarity between the predicted clutter-free image and the ground truth, representing a better clutter removal performance.\par

The comparisons of different methods using the four evaluation metrics on the simulated testing data are listed in Table \ref{table:comp_sim}. The metric values are the averaged results of all testing data. As shown in Table \ref{table:comp_sim}, the CR-Net outperforms other methods by a large margin under each evaluation metric. CAE and RPCA achieve the second and third best results, respectively. The conventional method SVD performs the worst. \par

\begin{table}[tb]
	\caption{Quantitative Comparisons of the Performance of Different Clutter Removal Methods. (The Best Results are Highlighted in \textcolor{blue}{Blue}.)}
	\centering
	\begin{tabular}{c c c c c}
		\hline
		\hline
		\textbf{Methods}   & \textbf{SVD} & \textbf{RPCA} & \textbf{CAE} & \textbf{CR-Net (ours)}  \\
		\hline
\textbf{MAE×10$^4$ (↓)}  &  59.62    &	 47.82   &	 29.62   &	 \textcolor{blue}{7.59}\\
\textbf{MSE×10$^4$ (↓)}  &  1.79    &	 1.40   &	 0.38   &	 \textcolor{blue}{0.04}\\
\textbf{PSNR (↑)}  &  39.82    &	 41.50   &	 45.77   &	 \textcolor{blue}{54.92}\\
\textbf{MS-SSIM (↑)}  &  0.9515    &	 0.9638   &	 0.9907   &	 \textcolor{blue}{0.9990}\\
		\hline
		\hline
	\end{tabular}
	\label{table:comp_sim}
\end{table}

\begin{figure}[t]
	\centering
	\centerline{\includegraphics[width=1\linewidth]{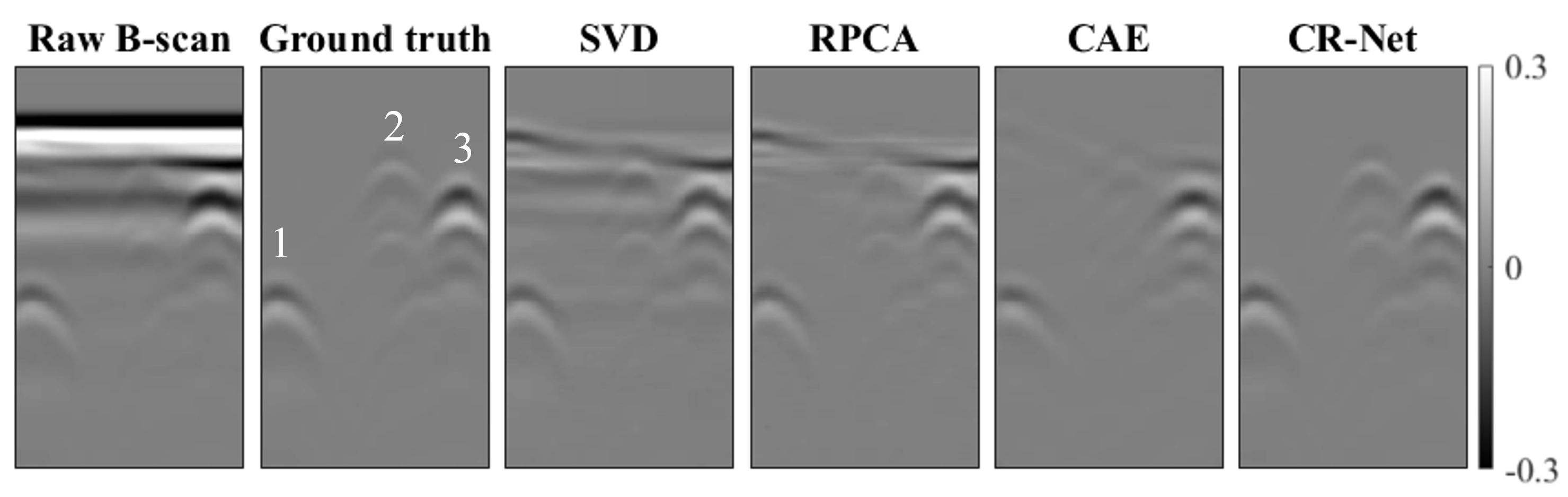}}
	\caption{Visual comparison of the clutter removal performance of different methods on the simulated testing data. The simulated scenario is three objects (1, 2, and 3 shown in ground truth) at different positions in heterogeneous soil with rough surface and water puddles. Compared with the ground truth, the CR-Net produces the best clutter-free radargram with indistinguishable clutter and intact object responses. (The radargrams are normalized into the range of [-0.5 0.5] for illustration.) }
	\label{fig:comp_sim}
\end{figure}

\begin{figure}[t]
	\begin{center}
		\begin{tabular}{c@{ }}
			\includegraphics[width=0.9\linewidth]{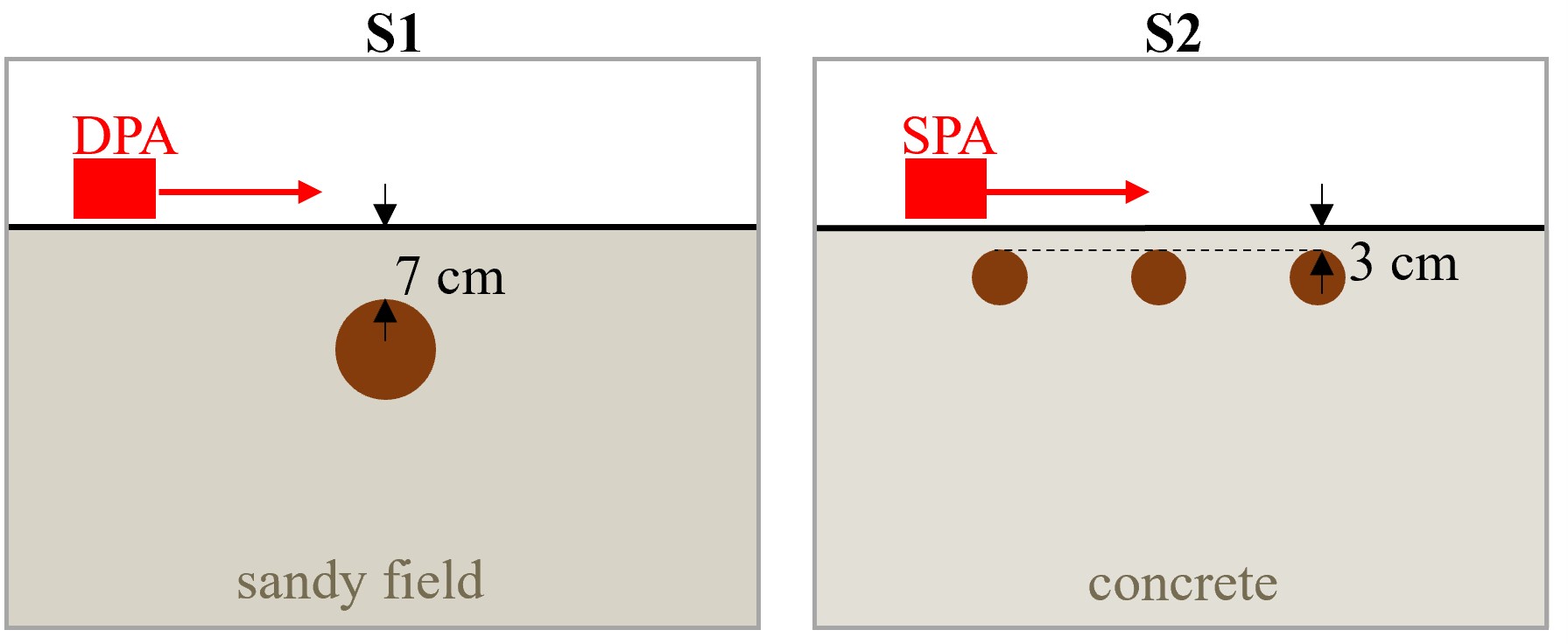}\\
			\footnotesize{(a)} \\
			\includegraphics[width=1\linewidth]{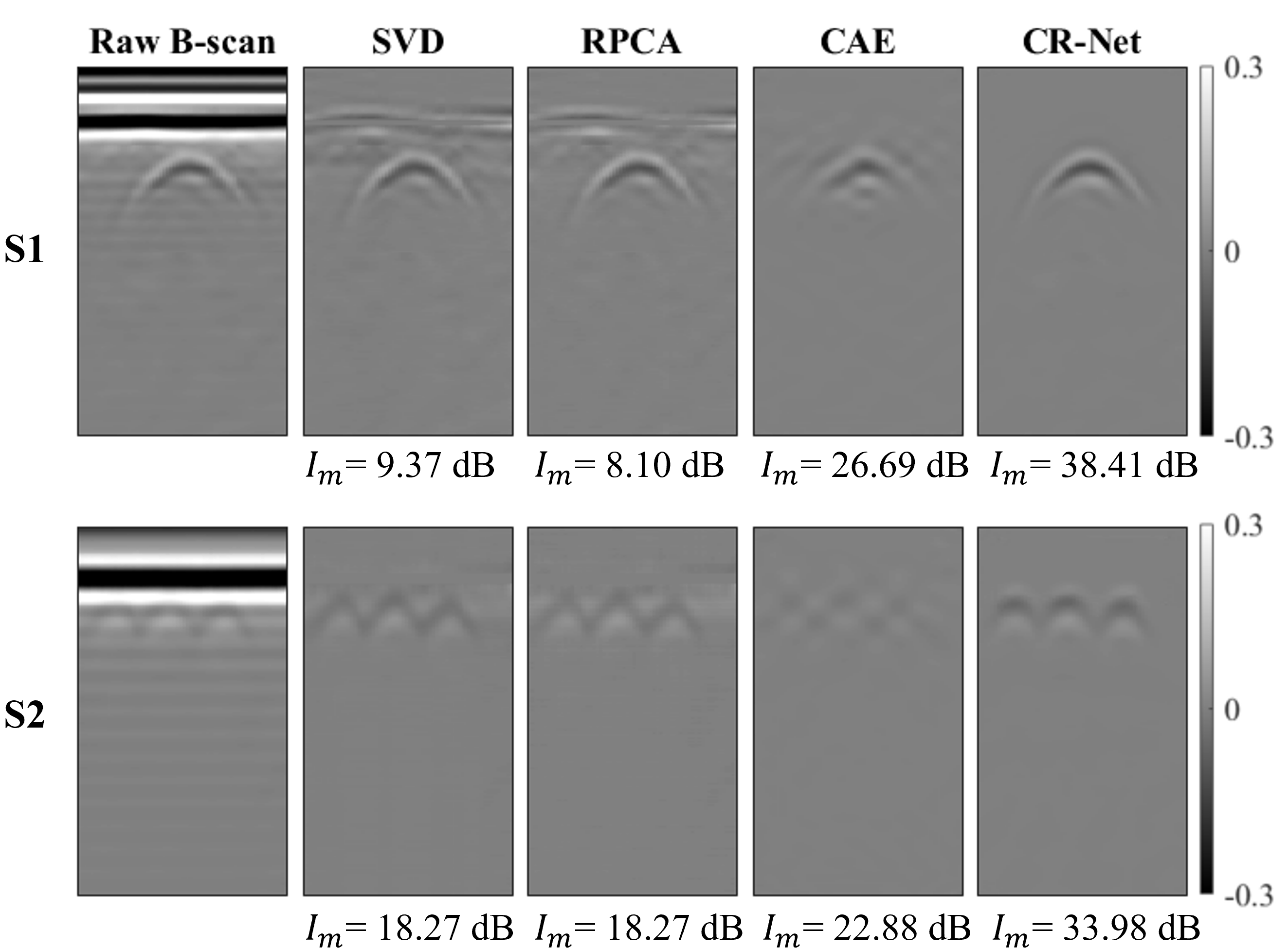}\\
			\footnotesize{(b)}  \\
		\end{tabular}
	\end{center}
	\caption{(a) Illustration of the experimental scenarios S1 and S2. (b) Comparison of the clutter removal performance of different methods on the measured radargrams. The improvement factor $I_m$ of different methods are also presented. Compared with other methods, the CR-Net successfully removes most of the clutter while restoring object reflections, and it achieves the highest $I_m$ values in the two cases.}
	\label{fig:comp_mea}
\end{figure}

\begin{figure*}[!tb]
	\centering
	\centerline{\includegraphics[width = 1\linewidth]{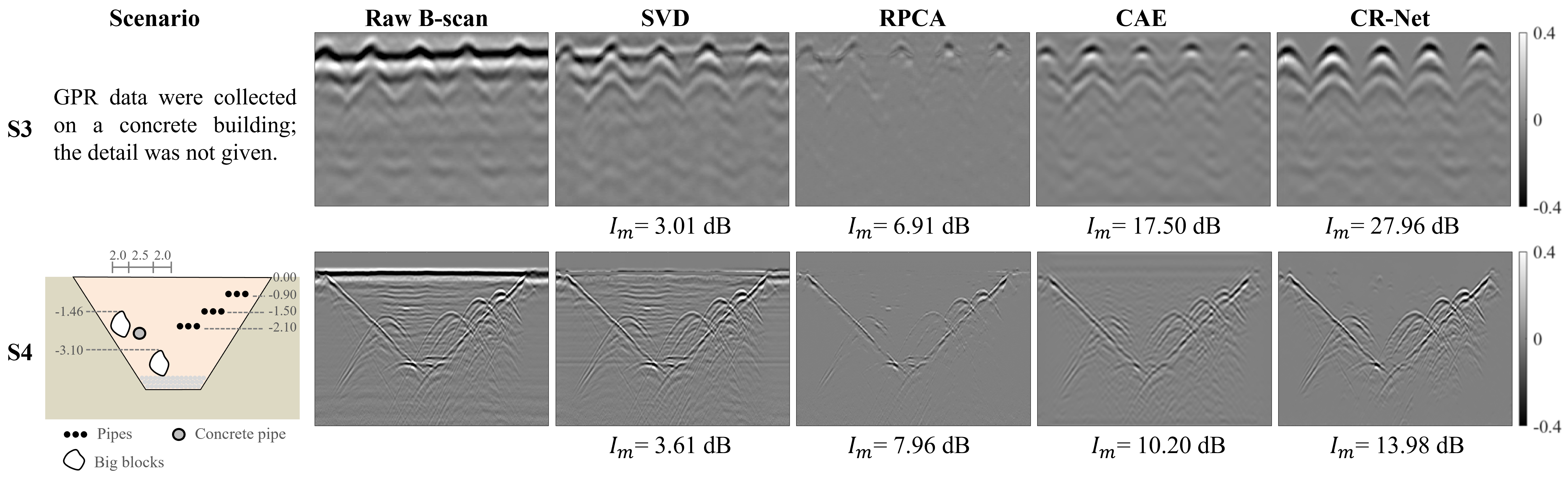}}
	\caption{Clutter removal performance of different methods on radargrams from open-source datasets. The radargram of scenario 3 (S3) and scenario 4 (S4) are from the rebar dataset \cite{rebar}, and the TU1208 dataset  \cite{tu1208}, respectively. In S3, the data were collected on a concrete building using the handy search NJJ-105 GPR. In S4, the 200-MHz GSSI system was used to survey a gneiss site with multiple subsurface objects. The $I_m$ values of the results are also presented. The CR-Net achieves the best performance in eliminating clutter and restoring  target responses.}
	\label{fig:comp_opensource}
\end{figure*}

The visual results of the CR-Net and other comparison methods on the simulated testing data are presented in Fig. \ref{fig:comp_sim}. One example of the most challenging scenario is examined where three objects are at different positions in heterogeneous soil with a rough surface and water puddles. The marked 1, 2, and 3 in the ground truth are the reflection signatures of the three subsurface objects. Compared with the ground truth, SVD and RPCA cannot fully remove the clutter caused by the uneven surface, water puddles, and soil heterogeneity, leaving pronounced clutter that interfered with reflection signature of the shallowly buried targets 2 and 3. CAE removes most of the clutter, but it also weakens the strength of the reflected signal of object 2, which is unwanted in the clutter removal process. The CR-Net, on the other hand, not only eliminates clutter but also well restores the target responses, producing the most promising clutter-free image. The visual comparison together with the quantitative assessment in Table \ref{table:comp_sim} demonstrates the superior performance of the CR-Net in removing clutter and restoring target responses. \par

\subsubsection{Results on Measured Data in Different Scenarios}
Since the ground-truth clutter-free images in the field experiments cannot be obtained, we use the improvement factor $I_m$ to quantitatively evaluate the clutter removal performance on measured radargrams. The improvement factor is defined as the ratio of the signal-to-clutter ratio of the processed radargram $ SCR_{processed}$ to the signal-to-clutter ratio of the raw radargram $ SCR_{raw}$ \cite{IM}:
\begin{equation}
\label{equ_12}
 I_m (dB)=20\cdot\log_{10}{\frac{{SCR}_{processed}}{{SCR}_{raw}}} .
\end{equation}
The signal-to-clutter ratio (SCR) is calculated as the ratio of the maximum amplitude of the target reflected signal $A_{max\_signal}$ to the maximum amplitude of the clutter $A_{max\_clutter}$ in the GPR B-scan:
\begin{equation}
\label{equ_13}
SCR=\frac{A_{max\_signal}}{A_{max\_clutter}}.
\end{equation}
A higher $I_m$ value corresponds to a better clutter removal performance.\par

The radargrams are collected in different scenarios using different GPR configurations, as illustrated in Fig. \ref{fig:comp_mea}(a). In scenario 1 (S1), an aluminum can is buried in a sandy field at a depth of 7 cm. The DPA as described in Section II.B is used to collect the object reflection. In scenario 2 (S2), three rebars are located at a cover depth of 3 cm in a concrete sample, and the SVA as described in Section II.C is employed to collect the rebar reflected signals. The raw B-scans in the two cases are shown in Fig. \ref{fig:comp_mea}(b). As the rebars in S2 are at shallow depths, a majority of their reflected signals are disguised by the strong surface clutter, which poses great challenges to extract object responses.\par

The clutter removal performance of different methods and their $I_m$ scores are compared in Fig. \ref{fig:comp_mea}(b). SVD and RPCA cannot completely suppress the background clutter, leaving residual clutter merged with object reflected signals, which is more obvious in the results of S2. CAE is effective in removing clutter, but it distorts or weakens parts of the target responses. The CR-Net, in contrast, successfully removes most of the clutter and well preserves the object responses. The high performance of CR-Net is more obvious in S2, where the interference of strong surface clutter on the reflection signatures of shallowly buried targets is significantly suppressed, and the target reflections are well extracted. The superior performance of the CR-Net can also be quantified using the $I_m$ values. The CR-Net achieves the highest $I_m$ values in both cases, whereas CAE remains as the second-best method and RPCA and SVD have the lowest $I_m$ values.\par

The comparison of results on the measured radargrams in this subsection shows that the CR-Net maintains its superior performance in removing real-world clutter and restoring object responses even when the object reflection is severely obscured by clutter.

\subsubsection{Results on the Open-Source GPR data}
The CR-Net is also implemented in open-source GPR radargrams to examine its generalization capability of removing clutter in completely new real-world scenarios. The raw radargrams of scenario 3 (S3) and scenario 4 (S4) are from the rebar dataset \cite{rebar}, and the TU1208 dataset \cite{tu1208}, respectively. In S3, the data were collected on a concrete building using the 1500-MHz NJJ-105 GPR. As shown in the raw B-scan of S3 in Fig. \ref{fig:comp_opensource}, the rebar reflection is mixed with a horizontally interfering signal, which challenges the rebar recognition and characterization \cite{rebar}. In S4, the 200-MHz GSSI system was used to survey a site filled with gneiss as depicted in Fig. \ref{fig:comp_opensource}. The site mainly consists of two dolmens, an empty concrete pipe, and three layers of pipes with an empty steel pipe, a PVC pipe full of water, and an empty PVC pipe per layer \cite{tu1208}. \par

The clutter removal performance of the CR-Net and other methods for the two scenarios is compared in Fig. \ref{fig:comp_opensource}. The $I_m$ scores of different methods are calculated and shown in the figure. In S3, SVD cannot fully remove the horizontal bandings and it affects the rebar reflected signals. This is because the clutter and rebar signals have similar strength in the raw data, so they cannot be well separated using different singular values.  RPCA removes most of the clutter, but it also severely distorts the rebar reflection. CAE and CR-Net both mitigate the clutter, but CR-Net achieves better performance in restoring the rebar reflections. In S4, the clutter removal capability of SVD is the worst with pronounced residual clutter and additional horizontal bandings. RPCA and CAE suppress most of the clutter, but they result in incomplete or weakened target responses. The CR-Net achieves the best performance in removing most of the clutter while keeping the reflection signatures of subsurface objects intact. The superior performance of the CR-Net is also quantified by the highest $I_m$ scores in both scenarios.\par

It is noted that the S3 and S4 are completely new scenarios, and the GPR systems employed are different from any GPR systems we have experimented with. Therefore, the clutter of S3 and S4 has different characteristics from our measured clutter. Inspiritingly, the CR-Net maintains its effectiveness in removing background clutter and restoring target responses. The results shown in this subsection demonstrate the good generalization capability of the CR-Net in eliminating clutter in various real-world scenarios.\par

\subsection{Comparison of Clutter Removal Performance of Different Network Architectures}
\begin{figure}[t]
	\centering
	\centerline{\includegraphics[width=1\linewidth]{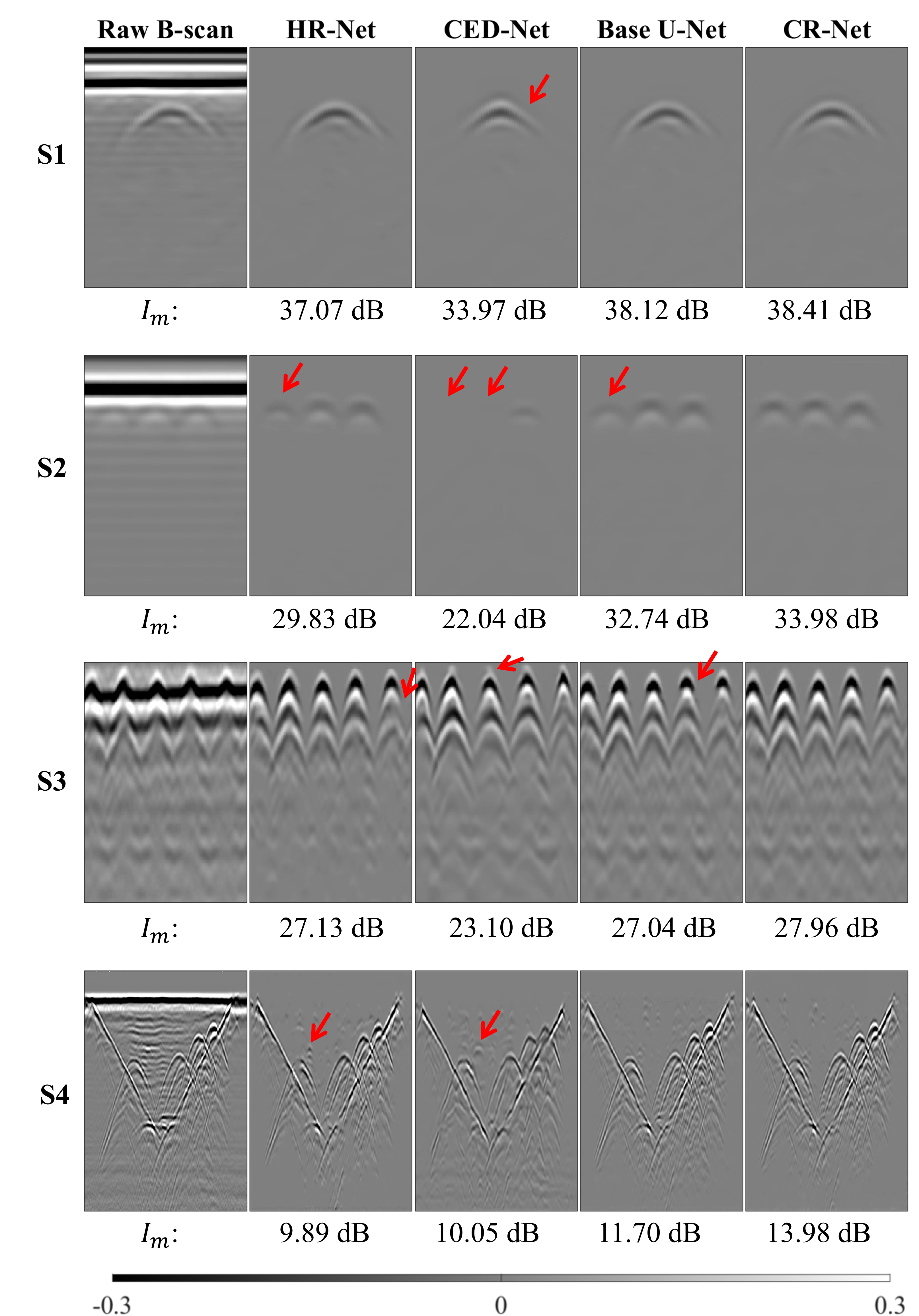}}
	\caption{Visual comparison of the clutter removal results of different network models. The improvement factor $I_m$ values of different methods are also presented. The HR-Net \cite{HRNet}, CAE-Net and base U-Net \cite{unet} leave residual clutter or distort parts of the target responses in some cases, as indicated by the red arrows. The final CR-Net achieves the best performance in removing clutter and restoring target responses. }
	\label{fig:comparenetwork}
\end{figure}

After demonstrating the superior clutter removal performance of the CR-Net over the conventional methods, its performance is also compared with that of different baseline deep learning network architectures. The compared network architectures include the high-resolution network (HR-Net) \cite{HRNet}, the convolutional encoder-decoder network (CED-Net), and the base U-Net \cite{unet}. For a fair comparison, we keep a similar amount of computational cost of each network. Specifically, the HR-Net is built using only the high-resolution convolution stream and the CED-Net is configured as the encoder-decoder structure in the base U-Net without the skip connections. The compared network architectures are trained using the same training settings, loss function, and training dataset as the proposed CR-Net. Their performance on both the synthetic testing data and the measured data is quantitatively and qualitatively compared in Table \ref{table:comparenetwork} and Fig. \ref{fig:comparenetwork}, respectively.\par

It can be seen from Table \ref{table:comparenetwork} and Fig. \ref{fig:comparenetwork} that the base U-Net achieves better clutter removal performance than the HR-Net and the CED-Net on both the synthetic testing data and the measured data. This is because the skip connections in the U-Net preserve more complete information of the target response during the clutter removal process.  Our proposed CR-Net performs the best in removing clutter and restoring target responses quantitatively and visually, as the addition of residual dense blocks adaptively preserves the features related to the target response and reduces the unwanted clutter features in the skip connection paths. The comparisons demonstrate the superiority of the base U-Net architecture and the further performance enhancement of our final CR-Net.\par

\begin{table}[tb]
	\caption{Comparison of Clutter Removal Performance of Different Network Architectures. (The Best Results are Highlighted in \textcolor{blue}{Blue}.)}
	\centering
	\begin{tabular}{p{1.7cm}<{\centering} p{1.1cm}<{\centering} p{1.2cm}<{\centering} p{1.5cm}<{\centering} p{1.1cm}<{\centering}}
		\hline
		\hline
		\textbf{Network architectures}   & \textbf{HR-Net \cite{HRNet}} & \textbf{CED-Net} & \textbf{Base U-Net \cite{unet}} & \textbf{CR-Net (ours)}  \\
		\hline
\textbf{MAE×10$^4$ (↓)}  &  11.93    &	 13.34   &	 9.01   &	 \textcolor{blue}{7.59}\\
\textbf{MSE×10$^4$ (↓)}  &  0.06    &	 0.09   &	 0.04   &	 \textcolor{blue}{0.04}\\
\textbf{PSNR (↑)}  &  52.75    &	 51.98   &	 54.21   &	 \textcolor{blue}{54.92}\\
\textbf{MS-SSIM (↑)}  &  0.9983    &	 0.9979   &	 0.9989   &	 \textcolor{blue}{0.9990}\\
		\hline
		\hline
	\end{tabular}
	\label{table:comparenetwork}
\end{table}

\subsection{Ablation Study}
\begin{figure*}[!htp]
	\centering
	\centerline{\includegraphics[width = 1\linewidth]{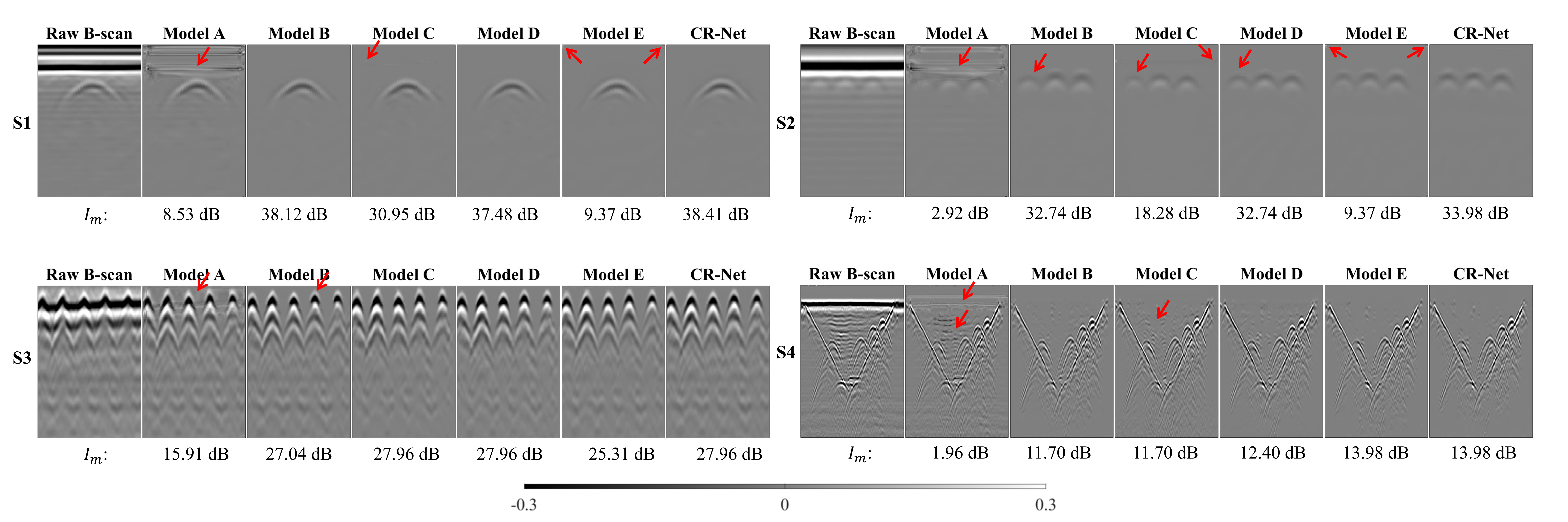}}
	\caption{Visual comparison of the clutter removal results of ablation models. The ablated models leave unremoved clutter, produce noticeable artifact, or distort parts of the target responses in some cases, as indicated by the red arrows. The final CR-Net achieves the best clutter removal capability with the clearest background and well-restored target responses. (Zoom in for a better view.)} 
	\label{fig:ablation}
\end{figure*}

Ablation studies are conducted to demonstrate the effectiveness of the main parts of our network design and implementation. The effects of the inclusion of hybrid sub-datasets into the training data, the integration of RDBs in the network architecture, and the utilization of different losses are investigated. The ablated models are retrained while keeping the same settings as the final model CR-Net except for the ablated parts. The performance of the ablated models on the synthetic testing data is quantitively compared in Table \ref{table:ablation}, and the performance on the measured data is visually compared in Fig. \ref{fig:ablation}.\par

\begin{table*}[!t]
	\caption{Comparison of Clutter Removal Performance in Ablation Study (The Best Results are Highlighted in \textcolor{blue}{Blue}.)}
	\centering
	\begin{tabular}{c|c|c|c  c  c| c}
		\hline
		\hline
		\textbf{ }   & \textbf{Effect of hybrid data} & \textbf{Effect of RDB} & \multicolumn{3}{|c|}{\textbf{Effects of different losses}} & \textbf{Final model}  \\
		\hline
\textbf{Models}  &  \textbf{A}    &	 \textbf{B}   &	 \textbf{C}   &	\textbf{D}  &	\textbf{E} & \textbf{CR-Net}\\
\hline
\textbf{Hybrid sub-datasets}  &     &	 \checkmark   &	 \checkmark   &	\checkmark  &	\checkmark & \checkmark\\
\textbf{RDB}  &  \checkmark   &	   &	 \checkmark  &	\checkmark  &	\checkmark & \checkmark\\
\textbf{MSE loss}  &                &	              & \checkmark	  &	  &	&  \\
\textbf{MAE loss}  &  \checkmark    &	 \checkmark   &	    & \checkmark	  &	 & \checkmark\\
\textbf{MS-SSIM loss}  &  \checkmark  &	 \checkmark   &	    &	  &	\checkmark & \checkmark\\
\hline
\textbf{MAE×10$^4$ (↓)}  &  9.89   &	 9.01   &	 15.93  &	10.11 &	8.66 & \textcolor{blue}{7.59}\\
\textbf{MSE×10$^4$ (↓)}  & 0.08   &	 0.04   &	 0.09  &	0.05  &	0.09 & \textcolor{blue}{0.04}\\
\textbf{PSNR (↑)}  &  51.37   &	 54.21   &	50.89  &	53.57 &	51.34 & \textcolor{blue}{54.92}\\
\textbf{MS-SSIM (↑)}  &  0.9985   &	0.9989   &	 0.9973   &	0.9986  &	0.9987 & \textcolor{blue}{0.9990}\\
		\hline
		\hline
	\end{tabular}
	\label{table:ablation}
\end{table*}

\textbf{Effect of adding hybrid sub-datasets into training data.} Model A is the network trained only on the simulated dataset. Comparing the performance of Model A and the CR-Net trained on both the simulated and hybrid datasets in Fig. \ref{fig:ablation}, it can be observed that the CR-Net exhibits a significant improvement in removing clutter in real-world GPR radargrams. Using only the simulated dataset as training data cannot allow the network to learn complex and diverse distributions of real clutter, resulting in a limited clutter removal capability of Model A. This is evident in Fig. \ref{fig:ablation} where the results of Model A have obvious unremoved clutter and artifacts. The hybrid sub-datasets built with clutter data collected by different GPR systems in multiple scenarios specifically addresses this issue. Adding hybrid datasets into training data enables the network to learn the random and complex distributions of real clutter, thereby greatly enhancing the clutter removal performance of the network in real-world radargrams. 

\textbf{Effect of integrating RDBs in the network architecture.} Model B is the base U-Net \cite{unet} without the integration of RDBs. As listed in Table \ref{table:ablation}, the performance of Model B falls behind the CR-Net on the simulated testing data. As shown in Fig. \ref{fig:ablation}, although Model B successfully removes most of the clutter, it distorts the target responses in S2 and S3. This could be due to the interference of clutter information from the skip-connected encoder features. Integrating RDBs into the skip connection paths helps to filter out the unwanted clutter information while preserving target responses in the concatenated features, thereby improving the clutter removal capability of the network, as demonstrated in the results of the CR-Net shown in Fig. \ref{fig:ablation}.\par

\textbf{Effect of using different losses.} Although the MSE loss is the dominant loss in most deep learning-based methods, it does not appear to be the best loss in the GPR clutter removal task. This is because MSE penalizes large errors but is more tolerant of small errors \cite{lossfunct}, which could not drive the network to obtain a complete restoration of the target response. In the ablation study, the performance of the network trained using three different losses is compared. Model C-E are the network trained using the MSE loss, MAE loss, and MS-SSIM loss, respectively. As shown in Table \ref{table:ablation}, Model C trained using the MSE loss has the worst score in each evaluation metric compared with Models D and E, showing the disadvantage of the MSE loss in the clutter removal task. As shown in Fig. \ref{fig:ablation}, Model C either leaves residual clutter or deteriorates target responses in most cases, which does not produce favorable clutter-free results. In contrast, Models D and E produce cleaner radargram and better preserve target signals in most scenarios. This comparison result suggests that the MAE loss and MS-SSIM loss can be more suitable loss functions than the MSE loss for the clutter removal task. However, although outperform the MSE loss, they still leave slight clutter or distort target reflections in some cases. The loss function combining the MAE loss and the MS-SSIM loss leverages the advantages of the two loss functions. It drives the network to achieve consistent and superior clutter removal performance in different scenarios, as demonstrated quantitatively and qualitatively in Table \ref{table:ablation} and Fig. \ref{fig:ablation}.\par

The ablation study proves that by integrating RDBs into the network architecture, training the network using the CLT-GPR dataset, and employing the combination of MAE and MS-SSIM losses to drive network optimization, the CR-Net enjoys superior and robust clutter removal capability in diverse scenarios. 

\subsection{Generalization Study}
After demonstrating the superior performance of the CR-Net for clutter removal in different environments using both simulated and measured data, additional investigations are conducted to verify the effectiveness of the CR-Net in a wider range of scenarios. These scenarios include the no-target case, the multi-layer case, the landmine case, and the GPR survey using different antennas, operating frequencies, and antenna-to-ground distances. \par

\begin{figure}[t]
	\centering
	\centerline{\includegraphics[width=1\linewidth]{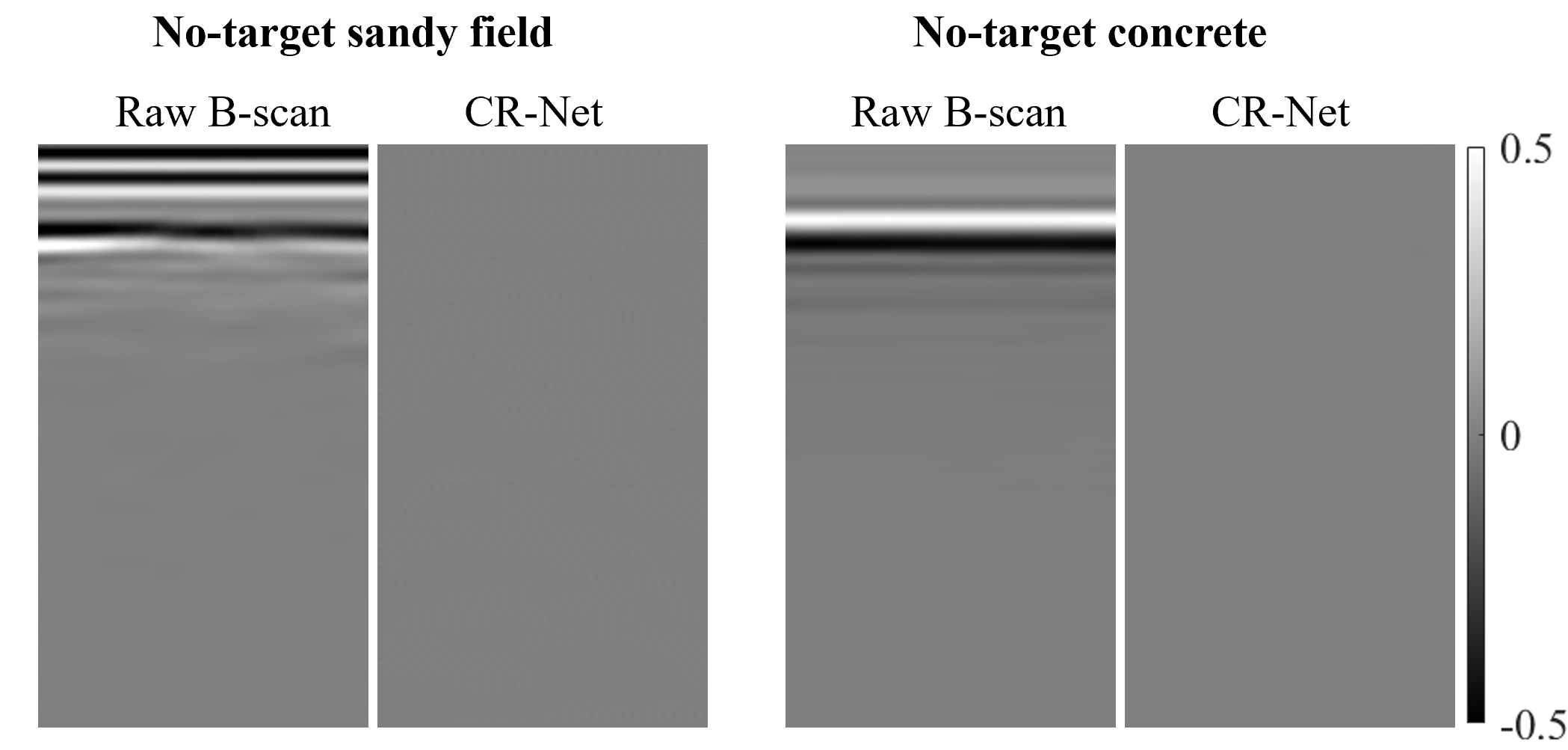}}
	\caption{The clutter removal performance of the CR-Net in no-target cases. The raw data are collected in the uneven sandy field and the concrete environment as described in Sections II.B and II.C. The CR-Net successfully removes environmental clutter without producing false targets and artifacts. }
	\label{fig:no-target}
\end{figure}

\begin{figure}[t]
	\centering
	\centerline{\includegraphics[width=1\linewidth]{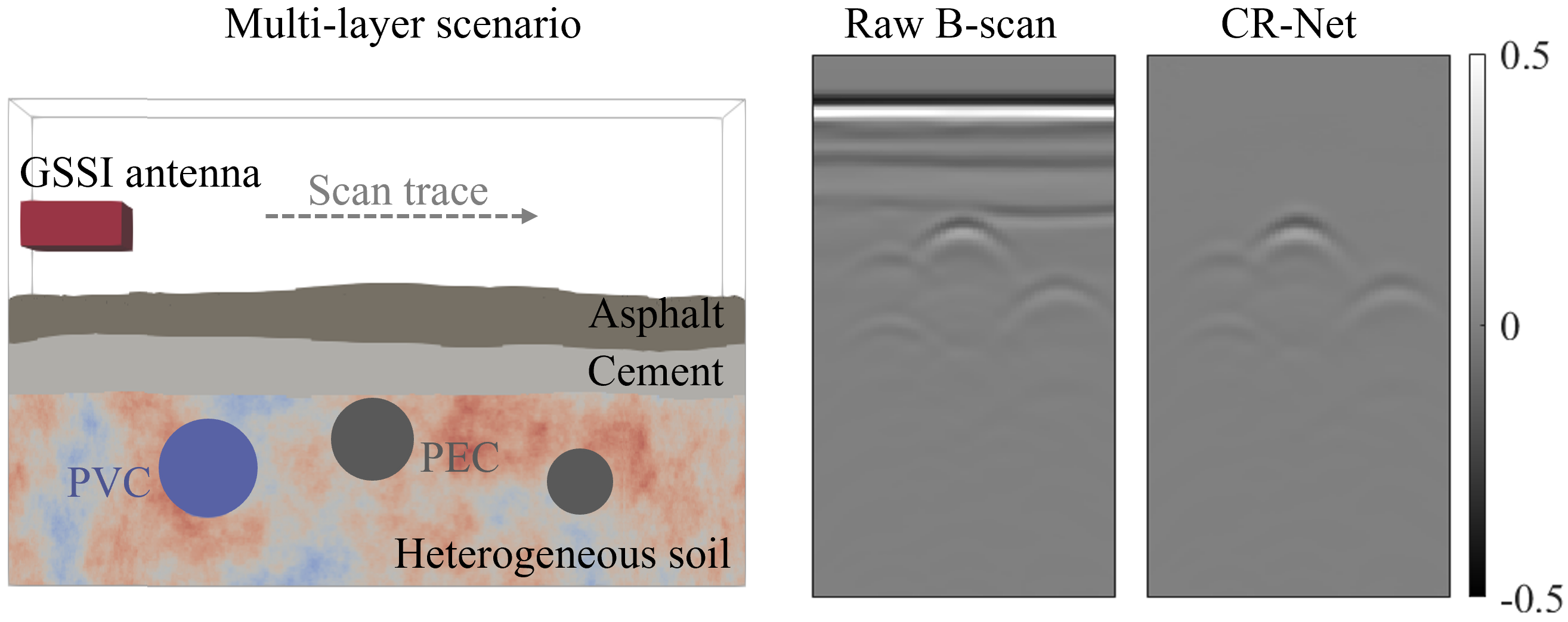}}
	\caption{The clutter removal performance of the CR-Net in the multi-layer scenario. The CR-Net removes the surface reflection at layer interfaces and well preserves the reflection of elongated targets, even when the interface between layers is rough and uneven. }
	\label{fig:multi-layer}
\end{figure}

\begin{figure}[th]
	\centering
	\centerline{\includegraphics[width=1\linewidth]{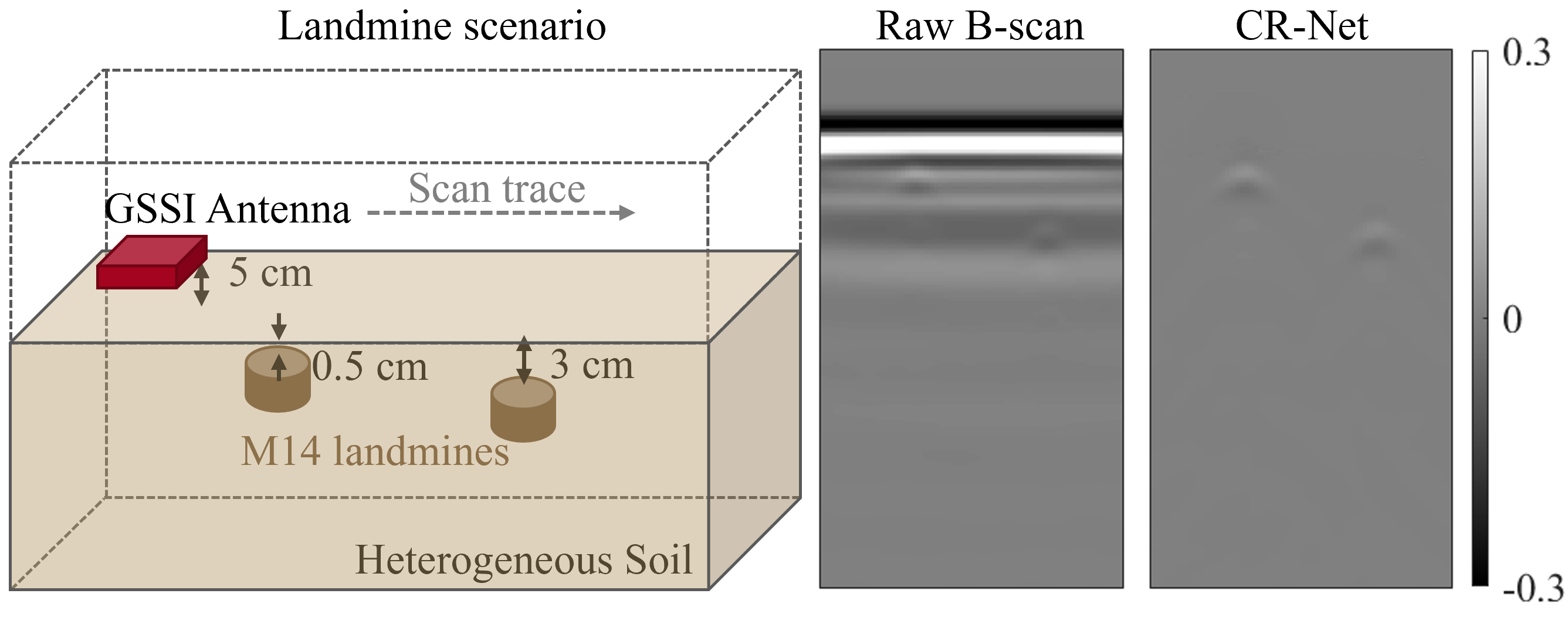}}
	\caption{The clutter removal performance of the CR-Net in the scenario with shallowly buried landmines. The CR-Net removes the environmental clutter and restores the reflection of the landmines.}
	\label{fig:landmine}
\end{figure}

\textbf{No-target case.} The measured data in the uneven sandy field and the concrete environment without any subsurface targets as described in Sections II.B and II.C are used to examine the performance of the CR-Net in the no-target cases. The raw radargrams and the processed radargrams by the CR-Net are shown in Fig. \ref{fig:no-target}. It can be seen that the CR-Net successfully removes the ground reflection and subsurface environmental clutter without producing false targets and artifacts.\par

\textbf{Multi-layer case.} The performance of the CR-Net is examined in a multi-layer scenario shown in Fig. \ref{fig:multi-layer}. The scenario consists of a 5-cm-thick asphalt layer with a relative permittivity of 5 and a conductivity of 0.001, a 5-cm-thick cement layer with a relative permittivity of 7 and a conductivity of 0.001, and a 20-cm thick heterogeneous soil layer, which is modeled to simulate the urban road environment \cite{rae}. Each layer has a rough surface with a height variation of 4 cm. One PVC pipe and two PEC pipes are buried within the depth range of 10-25 cm. The GPR survey setup is the same as described in Section II.A. The clutter removal performance of the CR-Net in the multi-layer scenario is shown in Fig. \ref{fig:multi-layer}. The CR-Net removes the surface reflection at layer interfaces and well preserves the reflection of elongated targets, even when the interfaces between layers are rough and uneven. The results demonstrate the applicability of the CR-Net in removing clutter and restoring target reflections in the multi-layer scenario.\par

\begin{figure*}[t]
	\centering
	\centerline{\includegraphics[width=1\linewidth]{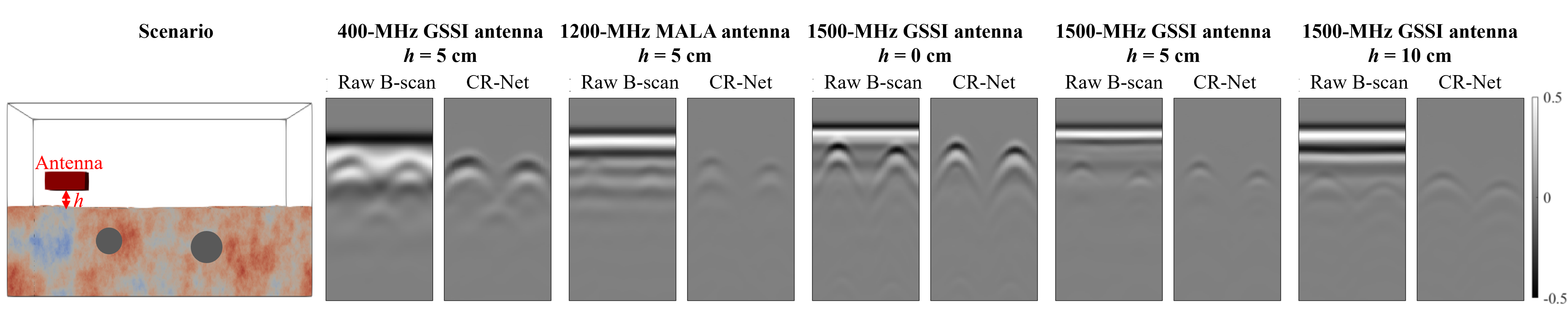}}
	\caption{The clutter removal performance of the CR-Net in the GPR survey using different antennas, operating frequencies, and antenna-to-ground distances. The CR-Net effectively and consistently removes environmental clutter in different cases. }
	\label{fig:diffantenna}
\end{figure*}

\textbf{Landmine case.} The clutter removal performance of the CR-Net is examined in a scenario with shallowly buried landmines as shown in Fig. \ref{fig:landmine}. Two cylindrical objects with a height of 4 cm, a diameter of 5.6 cm, a relative permittivity of 3.0 and a conductivity of 0.01 are modeled to simulate the M14 plastic-cased anti-personnel landmine \cite{cae}. They are buried in heterogeneous soil at shallow depths of 0.5 cm and 3 cm, respectively. The raw radargram and the processed radargram by the CR-Net are shown in Fig. \ref{fig:landmine}. The environmental clutter is successfully removed and the reflection signatures of the landmines are well restored. The results verify that the CR-Net maintains its effectiveness in clutter removal in scenarios with exposed targets such as landmines. \par

\textbf{GPR survey using different antennas, operating frequencies, and antenna-to-ground distances.} As GPR surveys use different antennas with different operating frequencies and have different antenna-to-ground distances in real-world applications, we further investigate the performance of the CR-Net in these cases. The subsurface scenario is the heterogeneous soil with a rough surface and two buried PEC pipes, as shown in Fig. \ref{fig:diffantenna}. Three different antennas operating at different frequencies are used as the transmitter and receiver. They are the 400-MHz GSSI antenna \cite{400mhzant}, the 1200-MHz MALA antenna \cite{gprmaxant1}, and the 1500-MHz GSSI antenna \cite{gprmaxant2}. For the 1500-MHz GSSI antenna, three different antenna-to-ground distances are modeled, which are 0 cm, 5 cm, and 10 cm, respectively. The raw B-scans and the processed B-scans by the CR-Net in different cases are shown in Fig. \ref{fig:diffantenna}. Although the clutter and the reflected signatures of the targets vary with the employed antennas, operating frequencies, and antenna-to-ground distances, the CR-Net removes the clutter and restores the object reflection with consistently high performance. It is noted that the 400-MHz GSSI antenna and the 1200-MHz MALA antenna are not used in generating the training dataset, yet the CR-Net maintains its effectiveness in these cases.  Therefore, the results demonstrate the versatility of the CR-Net in processing the radargrams collected by different antennas at different frequencies and with different antenna-to-ground distances. \par

The performance of the CR-Net in diverse scenarios illustrated in this subsection, together with the experimental results shown in Sections IV.A and IV.B, demonstrates that the performance of the CR-Net is not constraint by employed antennas, GPR operation frequencies, antenna-to-ground distances, and  subsurface environments. We will examine the performance of proposed method in more diverse scenarios in our future study.\par

\section{Conclusion}\label{sec5}
In this paper, a novel clutter removal method based on the neural network is presented. A network architecture, CR-Net, is designed to eliminate clutter and restore target responses in GPR B-scans. A large-scale dataset containing diverse and complex real-world clutter is built to train the network. The clutter diversity in the dataset greatly improves the generalization capability of the network in removing clutter in real-world radargrams. The effects of different loss functions on the network's clutter removal performance are explored, and a loss function that is more suitable for the clutter removal task is experimentally verified. Extensive experiments demonstrate that given real-world radargrams, the CR-Net can automatically and effectively remove clutter and restore target responses with consistently high performance. The proposed method outperforms the existing methods in  eliminating  clutter in diverse real-world scenarios and it does not require manual selection of hyperparameters.\par

The CR-Net with its robust clutter removal performance serves as a powerful tool to process real-world radargrams. It produces clutter-free images with well-restored target responses, which can facilitate further interpretation of radargrams such as target localization and characterization. \par

\end{document}